\documentclass[a4paper,12pt]{article}
\usepackage{epsfig}\parskip 5pt plus 1pt
\usepackage[english]{babel}
\usepackage{amsmath,amsfonts,amssymb,latexsym}
\usepackage{bbm}
\usepackage{cite}
\usepackage{amsmath}
\usepackage{amssymb}
\usepackage{amsfonts}
\usepackage{mathrsfs}
\usepackage{graphicx}
\usepackage{fancybox}
\usepackage{array}
\usepackage{axodraw}
\usepackage{multicol}
\usepackage{multirow}
\usepackage{lscape}
\usepackage{appendix}
\usepackage{pifont}
\usepackage{color}
\newcommand{\be}{\begin{equation}}
\newcommand{\ee}{\end{equation}}
\newcommand{\bea}{\begin{eqnarray}}
\newcommand{\eea}{\end{eqnarray}}

\newcommand{\bi}{\begin{itemize}}
\newcommand{\ei}{\end{itemize}}

\newcommand{\ie}{{\it i.e.}}

\newcommand{\eg}{{\it e.g.}}

\newcommand{\cf}{{\it cf.}}

\newcommand{\etc}{{\it etc.}}
\newcommand{\eq}{Eq.}

\newcommand{\fig}{Fig.}

\newcommand{\Ref}{Ref.}
\newcommand{\Refs}{Refs.}
\newcommand{\Sec}{Sec.}

\newcommand{\App}{the Appendix}

\newcommand{\Tab}{Table}

\newcommand{\equ}[1]{\eq~(\ref{equ:#1})}
\newcommand{\figu}[1]{\fig~\ref{Fig:#1}}

\newcommand{\FieldBoxOne[9]}{
$\varphi$ & $\phi'$ & $\psi$ & $\phi$\\
\cline{3-6}
\cline{3-6}
& &$1^{#2}_{#6}$ & $2^{#3}_{#7}$ & $1^{#4}_{#8}$ & $2^{#5}_{#9}$ \\
\cline{3-6}
& &$1^{#2}_{#6}$ & $2^{#3}_{#7}$ & $3^{#4}_{#8}$ & $2^{#5}_{#9}$\\
\cline{3-6}
& &$2^{#2}_{#6}$ & $1^{#3}_{#7}$ & $2^{#4}_{#8}$ & $1^{#5}_{#9}$\\
\cline{3-6}
& &$2^{#2}_{#6}$ & $1^{#3}_{#7}$ & $2^{#4}_{#8}$ & $3^{#5}_{#9}$\\
\cline{3-6}
& &$2^{#2}_{#6}$ & $3^{#3}_{#7}$ & $2^{#4}_{#8}$ & $1^{#5}_{#9}$\\
\cline{3-6}
& &$2^{#2}_{#6}$ & $3^{#3}_{#7}$ & $2^{#4}_{#8}$ & $3^{#5}_{#9}$\\
\cline{3-6}
& &$3^{#2}_{#6}$ & $2^{#3}_{#7}$ & $1^{#4}_{#8}$ & $2^{#5}_{#9}$\\
\cline{3-6}
& &$3^{#2}_{#6}$ & $2^{#3}_{#7}$ & $3^{#4}_{#8}$ & $2^{#5}_{#9}$
}

\newcommand{\FieldBoxTwo[9]}{
$\psi$ & $\phi$ & $\phi'$ & $\psi'$\\
\cline{3-6}
\cline{3-6}
& &$1^{#2}_{#6}$ & $2^{#3}_{#7}$ & $1^{#4}_{#8}$ & $2^{#5}_{#9}$ \\
\cline{3-6}
& &$1^{#2}_{#6}$ & $2^{#3}_{#7}$ & $3^{#4}_{#8}$ & $2^{#5}_{#9}$\\
\cline{3-6}
& &$2^{#2}_{#6}$ & $1^{#3}_{#7}$ & $2^{#4}_{#8}$ & $1^{#5}_{#9}$\\
\cline{3-6}
& &$2^{#2}_{#6}$ & $1^{#3}_{#7}$ & $2^{#4}_{#8}$ & $3^{#5}_{#9}$\\
\cline{3-6}
& &$2^{#2}_{#6}$ & $3^{#3}_{#7}$ & $2^{#4}_{#8}$ & $1^{#5}_{#9}$\\
\cline{3-6}
& &$2^{#2}_{#6}$ & $3^{#3}_{#7}$ & $2^{#4}_{#8}$ & $3^{#5}_{#9}$\\
\cline{3-6}
& &$3^{#2}_{#6}$ & $2^{#3}_{#7}$ & $1^{#4}_{#8}$ & $2^{#5}_{#9}$\\
\cline{3-6}
& &$3^{#2}_{#6}$ & $2^{#3}_{#7}$ & $3^{#4}_{#8}$ & $2^{#5}_{#9}$
}

\newcommand{\FieldBoxThree[9]}{
$\Psi$& $\psi'$ & $\phi$ & $\psi$\\
\cline{3-6}
\cline{3-6}
& &$1^{#2}_{#6}$ & $2^{#3}_{#7}$ & $1^{#4}_{#8}$ & $2^{#5}_{#9}$ \\
\cline{3-6}
& &$1^{#2}_{#6}$ & $2^{#3}_{#7}$ & $3^{#4}_{#8}$ & $2^{#5}_{#9}$\\
\cline{3-6}
& &$2^{#2}_{#6}$ & $1^{#3}_{#7}$ & $2^{#4}_{#8}$ & $1^{#5}_{#9}$\\
\cline{3-6}
& &$2^{#2}_{#6}$ & $1^{#3}_{#7}$ & $2^{#4}_{#8}$ & $3^{#5}_{#9}$\\
\cline{3-6}
& &$2^{#2}_{#6}$ & $3^{#3}_{#7}$ & $2^{#4}_{#8}$ & $1^{#5}_{#9}$\\
\cline{3-6}
& &$2^{#2}_{#6}$ & $3^{#3}_{#7}$ & $2^{#4}_{#8}$ & $3^{#5}_{#9}$\\
\cline{3-6}
& &$3^{#2}_{#6}$ & $2^{#3}_{#7}$ & $1^{#4}_{#8}$ & $2^{#5}_{#9}$\\
\cline{3-6}
& &$3^{#2}_{#6}$ & $2^{#3}_{#7}$ & $3^{#4}_{#8}$ & $2^{#5}_{#9}$
}

\newcommand{\FieldPyramid[7]}{ 
$\phi'$ & $\phi$ & $\psi$ \\
\cline{3-5}
\cline{3-5}
& &$1^{#2}_{#5}$ & $3^{#3}_{#6}$ & $2^{#4}_{#7}$ \\
\cline{3-5}
& &$2^{#2}_{#5}$ & $2^{#3}_{#6}$ & $1^{#4}_{#7}$ \\
\cline{3-5}
& &$2^{#2}_{#5}$ & $2^{#3}_{#6}$ & $3^{#4}_{#7}$ \\
\cline{3-5}
& &$3^{#2}_{#5}$ & $1^{#3}_{#6}$ & $2^{#4}_{#7}$ \\
\cline{3-5}
& &$3^{#2}_{#5}$ & $3^{#3}_{#6}$ & $2^{#4}_{#7}$ \\
\cline{3-5}
&&\multicolumn{3}{|c|}{}
}

\newcommand{\FieldTadpoleFive[7]}{
$\Delta$ & $\eta$ & $\eta'$\\
\cline{3-5}
\cline{3-5}
& &$3^{#2}_{#5}$ & $1^{#3}_{#6}$ & $3^{#4}_{#7}$ \\
\cline{3-5}
& &$3^{#2}_{#5}$ & $2^{#3}_{#6}$ & $2^{#4}_{#7}$ \\
\cline{3-5}
& &$3^{#2}_{#5}$ & $3^{#3}_{#6}$ & $1^{#4}_{#7}$\\
\cline{3-5}
&&\multicolumn{3}{|c|}{}\\ 
&&\multicolumn{3}{|c|}{}\\ 
&&\multicolumn{3}{|c|}{}\\
&&\multicolumn{3}{|c|}{}
}

\newcommand{\FieldTadpoleSix[7]}{
$\Delta$ & $\eta$ & $\eta'$\\
\cline{3-5}
\cline{3-5}
& &$3^{#2}_{#5}$ & $1^{#3}_{#6}$ & $2^{#4}_{#7}$ \\
\cline{3-5}
& &$3^{#2}_{#5}$ & $2^{#3}_{#6}$ & $1^{#4}_{#7}$ \\
\cline{3-5}
& &$3^{#2}_{#5}$ & $2^{#3}_{#6}$ & $3^{#4}_{#7}$\\
\cline{3-5}
& &$3^{#2}_{#5}$ & $3^{#3}_{#6}$ & $2^{#4}_{#7}$\\
\cline{3-5}
&&\multicolumn{3}{|c|}{}\\ 
&&\multicolumn{3}{|c|}{}\\ 
&&\multicolumn{3}{|c|}{}
}

\newcommand{\FieldTriangleTFourThreei[9]}{
$\Psi$ & $\phi$ & $\phi'$ & $\psi$\\
\cline{3-6}
\cline{3-6}
& &$1^{#2}_{#6}$ & $1^{#3}_{#7}$ & $1^{#4}_{#8}$ & $2^{#5}_{#9}$ \\
\cline{3-6}
& &$1^{#2}_{#6}$ & $2^{#3}_{#7}$ & $2^{#4}_{#8}$ & $1^{#5}_{#9}$\\
\cline{3-6}
& &$1^{#2}_{#6}$ & $2^{#3}_{#7}$ & $2^{#4}_{#8}$ & $3^{#5}_{#9}$\\
\cline{3-6}
& &$1^{#2}_{#6}$ & $3^{#3}_{#7}$ & $3^{#4}_{#8}$ & $2^{#5}_{#9}$\\
\cline{3-6}
& &$3^{#2}_{#6}$ & $1^{#3}_{#7}$ & $3^{#4}_{#8}$ & $2^{#5}_{#9}$\\
\cline{3-6}
& &$3^{#2}_{#6}$ & $2^{#3}_{#7}$ & $2^{#4}_{#8}$ & $1^{#5}_{#9}$\\
\cline{3-6}
& &$3^{#2}_{#6}$ & $2^{#3}_{#7}$ & $2^{#4}_{#8}$ & $3^{#5}_{#9}$\\
\cline{3-6}
& &$3^{#2}_{#6}$ & $3^{#3}_{#7}$ & $1^{#4}_{#8}$ & $2^{#5}_{#9}$\\
\cline{3-6}
& &$3^{#2}_{#6}$ & $3^{#3}_{#7}$ & $3^{#4}_{#8}$ & $2^{#5}_{#9}$
}

\newcommand{\FieldTriangleTFourThreeii[9]}{
$\Psi$ & $\psi$ & $\phi$ & $\psi'$\\
\cline{3-6}
\cline{3-6}
& &$1^{#2}_{#6}$ & $1^{#3}_{#7}$ & $1^{#4}_{#8}$ & $2^{#5}_{#9}$ \\
\cline{3-6}
& &$1^{#2}_{#6}$ & $2^{#3}_{#7}$ & $2^{#4}_{#8}$ & $1^{#5}_{#9}$\\
\cline{3-6}
& &$1^{#2}_{#6}$ & $2^{#3}_{#7}$ & $2^{#4}_{#8}$ & $3^{#5}_{#9}$\\
\cline{3-6}
& &$1^{#2}_{#6}$ & $3^{#3}_{#7}$ & $3^{#4}_{#8}$ & $2^{#5}_{#9}$\\
\cline{3-6}
& &$3^{#2}_{#6}$ & $1^{#3}_{#7}$ & $3^{#4}_{#8}$ & $2^{#5}_{#9}$\\
\cline{3-6}
& &$3^{#2}_{#6}$ & $2^{#3}_{#7}$ & $2^{#4}_{#8}$ & $1^{#5}_{#9}$\\
\cline{3-6}
& &$3^{#2}_{#6}$ & $2^{#3}_{#7}$ & $2^{#4}_{#8}$ & $3^{#5}_{#9}$\\
\cline{3-6}
& &$3^{#2}_{#6}$ & $3^{#3}_{#7}$ & $1^{#4}_{#8}$ & $2^{#5}_{#9}$\\
\cline{3-6}
& &$3^{#2}_{#6}$ & $3^{#3}_{#7}$ & $3^{#4}_{#8}$ & $2^{#5}_{#9}$
}

\newcommand{\FieldTriangleTFourOnei[9]}{
$\Delta$ & $\psi'$ & $\psi$ & $\Psi$\\
\cline{3-6}
\cline{3-6}
& &$3^{#2}_{#6}$ & $1^{#3}_{#7}$ & $3^{#4}_{#8}$ & $2^{#5}_{#9}$\\
\cline{3-6}
& &$3^{#2}_{#6}$ & $2^{#3}_{#7}$ & $2^{#4}_{#8}$ & $1^{#5}_{#9}$\\
\cline{3-6}
& &$3^{#2}_{#6}$ & $2^{#3}_{#7}$ & $2^{#4}_{#8}$ & $3^{#5}_{#9}$\\
\cline{3-6}
& &$3^{#2}_{#6}$ & $3^{#3}_{#7}$ & $1^{#4}_{#8}$ & $2^{#5}_{#9}$\\
\cline{3-6}
& &$3^{#2}_{#6}$ & $3^{#3}_{#7}$ & $3^{#4}_{#8}$ & $2^{#5}_{#9}$\\
\cline{3-6}
&&\multicolumn{4}{|c|}{}\\
&&\multicolumn{4}{|c|}{}
}

\newcommand{\FieldTriangleTFourOneii[9]}{
$\Delta$ & $\phi$ & $\phi'$ & $\Phi$\\
\cline{3-6}
\cline{3-6}
& &$3^{#2}_{#6}$ & $1^{#3}_{#7}$ & $3^{#4}_{#8}$ & $2^{#5}_{#9}$\\
\cline{3-6}
& &$3^{#2}_{#6}$ & $2^{#3}_{#7}$ & $2^{#4}_{#8}$ & $1^{#5}_{#9}$\\
\cline{3-6}
& &$3^{#2}_{#6}$ & $2^{#3}_{#7}$ & $2^{#4}_{#8}$ & $3^{#5}_{#9}$\\
\cline{3-6}
& &$3^{#2}_{#6}$ & $3^{#3}_{#7}$ & $1^{#4}_{#8}$ & $2^{#5}_{#9}$\\
\cline{3-6}
& &$3^{#2}_{#6}$ & $3^{#3}_{#7}$ & $3^{#4}_{#8}$ & $2^{#5}_{#9}$\\
\cline{3-6}
&&\multicolumn{4}{|c|}{}\\
&&\multicolumn{4}{|c|}{}
}

\newcommand{\FieldTriangleTFourTwoi[9]}{
$\Delta$ & $\phi'$ & $\phi$ & $\psi$\\
\cline{3-6}
\cline{3-6}
& &$3^{#2}_{#6}$ & $1^{#3}_{#7}$ & $3^{#4}_{#8}$ & $2^{#5}_{#9}$\\
\cline{3-6}
& &$3^{#2}_{#6}$ & $2^{#3}_{#7}$ & $2^{#4}_{#8}$ & $1^{#5}_{#9}$\\
\cline{3-6}
& &$3^{#2}_{#6}$ & $2^{#3}_{#7}$ & $2^{#4}_{#8}$ & $3^{#5}_{#9}$\\
\cline{3-6}
& &$3^{#2}_{#6}$ & $3^{#3}_{#7}$ & $1^{#4}_{#8}$ & $2^{#5}_{#9}$\\
\cline{3-6}
& &$3^{#2}_{#6}$ & $3^{#3}_{#7}$ & $3^{#4}_{#8}$ & $2^{#5}_{#9}$\\
\cline{3-6}
&&\multicolumn{4}{|c|}{}\\
&&\multicolumn{4}{|c|}{}
}

\newcommand{\FieldTriangleTFourTwoii[9]}{
$\Delta$ & $\psi'$ & $\psi$ & $\phi$\\
\cline{3-6}
\cline{3-6}
& &$3^{#2}_{#6}$ & $1^{#3}_{#7}$ & $3^{#4}_{#8}$ & $2^{#5}_{#9}$\\
\cline{3-6}
& &$3^{#2}_{#6}$ & $2^{#3}_{#7}$ & $2^{#4}_{#8}$ & $1^{#5}_{#9}$\\
\cline{3-6}
& &$3^{#2}_{#6}$ & $2^{#3}_{#7}$ & $2^{#4}_{#8}$ & $3^{#5}_{#9}$\\
\cline{3-6}
& &$3^{#2}_{#6}$ & $3^{#3}_{#7}$ & $1^{#4}_{#8}$ & $2^{#5}_{#9}$\\
\cline{3-6}
& &$3^{#2}_{#6}$ & $3^{#3}_{#7}$ & $3^{#4}_{#8}$ & $2^{#5}_{#9}$\\
\cline{3-6}
&&\multicolumn{4}{|c|}{}\\
&&\multicolumn{4}{|c|}{}
}

\newcommand{\mnuToneone}{\mu \mu' \left[\begin{array}{c} 
(Y^{\sf T})^{\alpha a} 
M_{a}(y)_{a}^{\beta}\\+
(y^{\sf T})^{\alpha}_{a}
M_{a} (Y)^{a \beta}\end{array}\right]
 I_{4} (M_{a}^{2}, m_{\phi}^{2}, m_{\phi'}^{2}, m_{\varphi}^{2})}

\newcommand{\mnuTonetwo}{\begin{array}{c}\mu\left[\begin{array}{c} 
(y'^{\sf T})^{\alpha}_{ a'}
M'_{a'} 
(Y_{L})_{a'}^{a}M_{a}(y)_{a}^{\beta}\\
+(y^{\sf T})^{\alpha}_{a}
M_{a} 
(Y^{\sf T}_{L})^{a}_{a'}M'_{a'}(y')^{a'}_{ \beta}\end{array}\right]
 I_{4} (M'^{2}_{a'}, M^{2}_{a},m_{\phi'}^{2},m_{\phi}^{2})\\
 \\
+ \mu
\left[\begin{array}{c}
(y'^{\sf T})^{\alpha a'}
(Y_{R})_{a'}^{a}(y)_{a}^{\beta}\\
+
(y^{\sf T})^{\alpha}_{a}
(Y^{\sf T}_{R})^{a}_{a'}(y')^{a' \beta}\end{array}
\right]  J_{4} (M'^{2}_{a'}, M^{2}_{a},m_{\phi'}^{2},m_{\phi}^{2})\end{array}}

\newcommand{\mnuTonethree}{\begin{array}{c}\left[\begin{array}{c}
(y'^{\sf T})^{\alpha a'}M'_{a'}
(Y'_{L})_{a'}^{A}
\mathcal{M}_{A}
(Y_{L})_{A}^{a}
M_{a}(y)_{a}^{\beta}
\\
+
(y^{\sf T})^{\alpha}_{a}
M_{a}
(Y_{L}^{\sf T})^{a}_{A}
\mathcal{M}_{A}
(Y'^{\sf T}_{L})^{A}_{a'}
M'_{a'}
(y')^{a' \alpha}\end{array}\right]\vspace{0.2cm}
I_{4} (M'^{2}_{a'}, M_{a}^{2}, \mathcal{M}_{A}^{2}, m_{\phi}^{2})\\

+
\left[\begin{array}{c}
(y'^{\sf T})^{\alpha a'}
M'_{a'}
(Y'_{L})_{a'}^{A}
(Y_{R})_{A}^{a}
(y)_{a}^{\beta}\\
+
(y^{\sf T})^{\alpha}_{a}
(Y_{R}^{\sf T})^{a}_{A}
(Y'^{\sf T}_{L})^{A}_{a'}
M'_{a'} 
(y')^{a' \beta}
\\
+
(y'^{\sf T})^{\alpha a'}
(Y'_{R})_{a'}^{A}
(Y_{L})_{A}^{a}
M_{a}
(y)_{a}^{\beta}\\
+
(y^{\sf T})^{\alpha}_{a}
M_{a}
(Y_{L}^{\sf T})^{a}_{A}
(Y'^{\sf T}_{R})^{A}_{a'}
(y')^{a' \beta}
\\
+
(y'^{\sf T})^{\alpha a'}
(Y'_{R})_{a'}^{A}
\mathcal{M}_{A}
(Y_{R})_{A}^{a}
(y)_{a}^{\beta}\\
+
(y^{\sf T})^{\alpha}_{a}
(Y^{\sf T}_{R})^{a}_{A}
\mathcal{M}_{A}
(Y'^{\sf T}_{R})^{A}_{a'}
(y')^{a' \beta}\end{array}\right]
 J_{4} (M'^{2}_{a'}, M_{a}^{2}, \mathcal{M}_{A}^{2}, m_{\phi}^{2})
\end{array}
}

\newcommand{\mnuTthree}{
- \lambda
\left[
\begin{array}{c}
(Y^{\sf T})^{\alpha a} 
M_{a}
{(y)_{a}}^{\beta}\\
+
{(y^{\sf T})^{\alpha}}_{a} 
M_{a}
(Y)^{a \beta}
\end{array}
\right] 
I_{3}(M_{a}^{2}, m_{\phi'}^{2}, m_{\phi}^{2})
}

\newcommand{\mnuTfouronei}{
4\frac{Y_{\Delta}^{\alpha\beta}}{m_{\Delta}^2}
\left\{
\begin{array}{cc}
&M_a M'_{a'} \mathcal{M}_A 
\left(
\begin{array}{c}
(y_L)^a_{a'} (Y_L)^A_a (Y'_L)^{a'}_A + L\leftrightarrow R 
\end{array}
\right)
I_3(M_a^2,{M'}_{a'}^2,\mathcal{M}_A^2)\\
&\\
+&\left[
\begin{array}{cc}
&M_a\left((y_L)^a_{a'} (Y_L)^A_a (Y'_R)^{a'}_A + L\leftrightarrow R \right)\\
+&M'_{a'}\left((y_L)^a_{a'} (Y_R)^A_a (Y'_L)^{a'}_A + L\leftrightarrow R \right)\\
+&\mathcal{M}_A\left((y_L)^a_{a'} (Y_R)^A_a (Y'_R)^{a'}_A + L\leftrightarrow R \right)
\end{array}
\right]
J_3(M_a^2,{M'}_{a'}^2,\mathcal{M}_A^2)
\end{array}
\right\}
}

\newcommand{\mnuTfouroneii}{
-4\frac{Y_{\Delta}^{\alpha\beta}}{m_{\Delta}^2}\mu_{\phi}\mu_{\Phi}\mu'_{\Phi}
I_3(m_\phi^2,{m}_{\phi'}^2,m_\Phi^2)
}

\newcommand{\mnuTfourtwoi}{
-\frac{\mu_{\Delta}}{m^2_{\Delta}}\mu_{\phi}
({y'}^T)^{\alpha a}M_a (y)_{a}^{\beta}
I_3(m_\phi^2,{m}_{\phi'}^2,M_a^2)
}

\newcommand{\mnuTfourtwoii}{
-\frac{\mu_{\Delta}}{m^2_{\Delta}}
\left\{
\begin{array}{cc}
&\left[
\begin{array}{cc}
&({y'}^T)^{\alpha a'}M'_{a'} (Y_L)_{a'}^a M_a (y)_a^{\beta}\\
+&({y}^T)^{\alpha}_a M_{a} (Y_L^T)_{a'}^a M'_{a'} (y')^{a' \beta}
\end{array}
\right]
I_3({M'}_{a'}^2,M_a^2,m_{\phi}^2)\\
&\\
+&\left[
\begin{array}{cc}
&({y'}^T)^{\alpha a'} (Y_R)_{a'}^a  (y)_a^{\beta}\\
+&({y}^T)^{\alpha}_a  (Y_R^T)_{a'}^a (y')^{a' \beta}
\end{array}
\right]
J_3({M'}_{a'}^2,M_a^2,m_{\phi}^2)
\end{array}
\right\}
}

\newcommand{\mnuTfourthreei}{
-\mu_{\phi}\left[\begin{array}{cc}
&
(Y_{\nu}^T)^{\alpha}_{\gamma}(M_{\Psi}^{-1})^{\gamma}
({y'})^{\gamma a} M_a (y)_{a}^{\beta} \\
+&({y}^T)^{\alpha}_{a} M_a (y'^{T})^{a \gamma} 
(M_{\Psi}^{-1})^{\gamma} (Y_{\nu})_{\gamma}^{\beta}
\end{array}
\right]
I_3(m_{\phi}^{2},m_{\phi'}^2,M_a^2)
}

\newcommand{\mnuTfourthreeii}{
\begin{array}{c}
-\left[\begin{array}{cc}
&
(Y_{\nu}^T)^{\alpha}_{\gamma}(M_{\Psi}^{-1})^{\gamma}(y^T)^{\gamma}_a (Y_L^T)^{a}_{a'} (y')^{a' \beta} \\
+&({y'}^T)^{\alpha a'} M'_{a'} (Y_L)_{a'}^a (y)_a^{\gamma} (M_{\Psi}^{-1})^{\gamma} (Y_{\nu})_{\gamma}^{\beta}
\end{array}
\right]M_a M'_{a'}
I_3(m_{\phi^2};M_a^2;{M'}_{a'}^2)\\
\\
-\left[\begin{array}{cc}
&
(Y_{\nu}^T)^{\alpha}_{\gamma}(M_{\Psi}^{-1})^{\gamma}(y^T)^{\gamma}_a (Y_R^T)^{a}_{a'} (y')^{a' \beta} \\
+&({y'}^T)^{\alpha a'} M'_{a'} (Y_R)_{a'}^a (y)_a^{\gamma} (M_{\Psi}^{-1})^{\gamma} (Y_{\nu})_{\gamma}^{\beta}
\end{array}
\right]
J_3(m_{\phi^2};M_a^2;{M'}_{a'}^2)
\end{array}
}

\newcommand{\mnuTfive}{
\frac{\lambda\mu Y_{\Delta}}{M_{\Delta}^2} I_2(M_{\eta}^{2},M_{\eta'}^{2})
}

\newcommand{\mnuTsix}{
\frac{\lambda\mu Y_{\Delta}}{M_{\Delta}^2} I_2(M_{\eta}^{2},M_{\eta'}^{2})
}

\begin{document}

\begin{titlepage}
\vspace{-1cm}
\begin{flushright}
\small
PREPRINT: IFIC/12-26, MPP-2012-76
\end{flushright}
\vspace{0.2cm}
\begin{center}
{\Large \bf Systematic study of the $\boldsymbol{d=5}$ Weinberg operator at one-loop order}
\vspace*{0.2cm}
\end{center}
\vskip0.2cm

\begin{center}
{\bf  Florian~Bonnet$^{a,}$\footnote{florian.bonnet@physik.uni-wuerzburg.de},  Martin~Hirsch$^{b,}$\footnote{mahirsch@ific.uv.es}, Toshihiko~Ota$^{c,}$\footnote{toshi@mppmu.mpg.de}, and Walter~Winter$^{a,}$\footnote{winter@physik.uni-wuerzburg.de}}
\end{center}
\vskip 8pt

\begin{center}
 {\it $^{a}$ Institut f\"{u}r Theoretische Physik und Astrophysik, Universit\"{a}t W\"{u}rzburg,\\
 97074 W\"{u}rzburg, Germany}  \\
{\it $^{b}$ AHEP Group, Instituto de F\'{\i}sica Corpuscular --
    C.S.I.C./Universitat de Val{\`e}ncia \\
    Edificio de Institutos de Paterna, Apartado 22085,
  46071 Val{\`e}ncia, Spain}\\
 {\it $^{c}$ Max-Planck-Institut f\"{u}r Physik (Werner-Heisenberg-Institut), F\"{o}hringer Ring 6\\
  80805 M\"{u}nchen, Germany}
\end{center}

\vspace*{0.3cm}

\vglue 0.3truecm

\begin{abstract}
\vskip 3pt \noindent

We perform a systematic study of the $d=5$ Weinberg operator at the
one-loop level. We identify three different categories of neutrino
mass generation: (1) finite irreducible diagrams; (2) finite
extensions of the usual seesaw mechanisms at one-loop and (3)
divergent loop realizations of the seesaws. All radiative one-loop
neutrino mass models must fall into one of these classes. Case (1)
gives the leading contribution to neutrino mass naturally and a
classic example of this class is the Zee model.  We demonstrate that
in order to prevent that a tree level contribution dominates in case
(2), Majorana fermions running in the loop and an additional
$\mathbb{Z}_2$ symmetry are needed for a genuinely leading one-loop
contribution. In the type-II loop extensions, the Yukawa coupling will be generated at one loop, whereas the
type-I/III extensions can be interpreted as loop-induced inverse
or linear seesaw mechanisms. For the divergent diagrams in category (3), 
the tree level contribution cannot be avoided and is in fact needed as
counter term to absorb the divergence.

\end{abstract}

\end{titlepage}

\newpage


\section{Introduction}


Neutrino oscillation experiments have firmly established the existence
of non-zero neutrino masses. The recent result of the Daya-Bay and Reno experiment 
give a non-zero value of $\theta_{13}$~\cite{An:2012eh,Ahn:2012nd} at more 
than 5 $\sigma$ C.L., finally fixing the last unknown leptonic mixing 
angle. This result has attracted a lot of attention, since it opens up 
the possibility to measure leptonic CP violation in the future.  With 
this data oscillation physics definitely has entered the precision age 
(for the latest fits of oscillation parameters see \eg\
\Refs~\cite{GonzalezGarcia:2010er,Schwetz:2011zk}).

Neutrino masses are at least six orders of magnitude smaller than 
the next lightest standard model fermion. If neutrinos are Majorana 
particles, such a small mass could be understood if there is new 
physics beyond the electroweak scale. The lowest order operator, 
which generates Majorana neutrino masses after EWSB, 
is the unique $d=5$ Weinberg operator~\cite{Weinberg:1979sa}:  
\begin{equation}
\delta\mathcal{L}=\frac{1}{2}c^{d=5}_{\alpha\beta}
\left(\overline{L^c_{L\alpha}}\widetilde{H}^*\right)
\left(\widetilde{H}^{\dagger}L_{L\beta}\right)+\rm{h.c.}\,,
\label{equ:d5}
\end{equation}
where $L_{L}=(\nu_L, \ell_L)^T$ are the usual left-handed lepton
doublets of the SM, $H=(H^+,H^0)^T$ is the Higgs doublet,
$\widetilde{H}=i \sigma_2 H^*$, and $c^{d=5}_{\alpha\beta} \propto
1/\Lambda$ is a model dependent coefficient suppressed by the scale of
new physics $\Lambda$. Here the Greek indices represent flavor
indices. All Majorana neutrino mass models reduce to this operator, or its higher dimensional ($d>5$) equivalent \cite{Bonnet:2009ej},
once the new physics is integrated out
\footnote{Note that current observational data, with a hint for the 
Higgs at $125 \,\mathrm{GeV}$~\cite{ATLAS:2012ae,Chatrchyan:2012tx}, 
is perfectly consistent with this picture.}.

\begin{figure}[htb]
\centering
\begin{picture}(400,110)
\DashLine(10,80)(40,50){4}
\ArrowLine(10,20)(40,50)
\ArrowLine(40,50)(60,50)
\ArrowLine(80,50)(60,50)
\DashLine(80,50)(110,80){4}
\ArrowLine(110,20)(80,50)
\Text(60,60)[]{$N$}
\Text(120,20)[]{$L$}
\Text(120,80)[]{$H$}
\Text(0,80)[]{$H$}
\Text(0,20)[]{$L$}
\Text(95,50)[]{$Y_{\nu_N}$}
\Text(25,50)[]{$Y_{\nu_N}^T$}
\DashLine(170,100)(200,70){4}
\ArrowLine(170,0)(200,30)
\DashLine(200,70)(200,30){4}
\DashLine(200,70)(230,100){4}
\ArrowLine(230,0)(200,30)
\Text(165,100)[]{$H$}
\Text(165,0)[]{$L$}
\Text(240,100)[]{$H$}
\Text(240,0)[]{$L$}
\Text(210,50)[]{$\Delta$}
\Text(200,85)[]{$\mu_\Delta$}
\Text(200,15)[]{$Y_\Delta$}
\DashLine(290,80)(320,50){4}
\ArrowLine(290,20)(320,50)
\ArrowLine(320,50)(340,50)
\ArrowLine(360,50)(340,50)
\DashLine(360,50)(390,80){4}
\ArrowLine(390,20)(360,50)
\Text(340,60)[]{$\Sigma$}
\Text(400,20)[]{$L$}
\Text(400,80)[]{$H$}
\Text(280,80)[]{$H$}
\Text(280,20)[]{$L$}
\Text(375,50)[]{$Y_{\nu_\Sigma}$}
\Text(305,50)[]{$Y_{\nu_\Sigma}^T$}
\end{picture}
\caption{\it The three generic realizations of the seesaw mechanism, 
depending on the nature of the heavy field exchanged: SM singlet fermion 
$N$ ($1^F_0$) (type~I seesaw) on the left, SM triplet scalars 
$\Delta$ ($3^S_{-2}$) (type~II seesaw) in the middle, and SM triplet 
fermion $\Sigma$ ($3^F_0$) (type~III seesaw) on the right.}
\label{Fig:TreeSeesaw}
\end{figure}
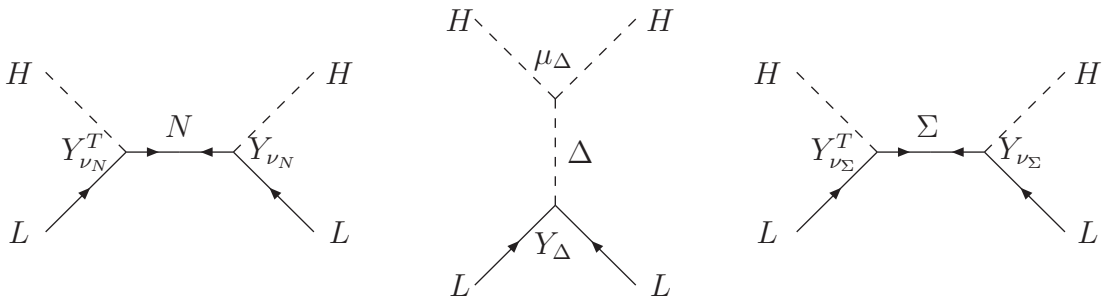

There are only three ways to generate the $d=5$ operator at tree level. 
These are known as type-I~\cite{Minkowski:1977sc,Yanagida:1979as,GellMann:1980vs,Mohapatra:1979ia},
type-II~\cite{Magg:1980ut,Schechter:1980gr,Wetterich:1981bx,Lazarides:1980nt,Mohapatra:1980yp,Cheng:1980qt},
and type-III\cite{Foot:1988aq} see-saw mechanisms, with an SU(2)
singlet fermion, triplet scalar, and triplet fermion, respectively, as
mediator; see also \Ref~\cite{Abada:2007ux} for a review. These three
possibilities are shown in \figu{TreeSeesaw}.

In the tree level seesaws, the neutrino mass scale is roughly given by
$m_\nu=\langle H^0 \rangle^2/\Lambda$, where $\langle H^0 \rangle$ is the vacuum expectation value (vev)
of the Higgs, as one can easily read off from \equ{d5}. For
$\mathcal{O}(1)$ couplings, $\Lambda$ points to the scale of a grand
unified theory (GUT), where the gauge couplings are supposedly
unifying. Therefore, in these standard seesaws, the corresponding new
physics scale will be impossible to test at the LHC. In turn, models
with a new physics scale naturally linked to physics at the LHC need
to have additional suppression mechanisms for neutrino mass, where the
best studied ones are:
\begin{enumerate}
\item
     The neutrino mass is generated radiatively. The additional suppression
     is guaranteed by a combination of loop 
     integrals~\cite{Zee:1980ai,Wolfenstein:1980sy,Zee:1985rj,Babu:1988ki,Ma:1998dn,FileviezPerez:2009ud,Choubey:2012ux,Babu:2002uu,Krauss:2002px,Cheung:2004xm,Ma:2006km,Ma:2007yx,Aoki:2008av,Aoki:2009vf,AristizabalSierra:2006gb} and EW-scale masses entering the diagrams. 
\item 
      The neutrino mass is generated at tree level, where additional
      suppression enters through a small lepton number violating (LNV)
      contribution (\eg, in inverse or linear see-saw scenarios, or  
      R parity-violating 
      SUSY models, \etc~\cite{Schechter:1981bd,Nandi:1985uh,Mohapatra:1986bd,Branco:1988ex,GonzalezGarcia:1988rw,Xing:2009hx,Ma:2000cc,Tully:2000kk,Loinaz:2003gc,Hirsch:2004he,Pilaftsis:2005rv,deGouvea:2007xp,Kersten:2007vk,Grimus:2009mm,Gavela:2009cd,Akhmedov:1995ip,Akhmedov:1995vm}).
\item
     The neutrino mass is forbidden or suppressed at $d=5$, but appears from 
     effective operators of higher dimension~\cite{Babu:1999me,Chen:2006hn,Gogoladze:2008wz,Giudice:2008uua,Babu:2009aq,Gu:2009hu,Bonnet:2009ej,Picek:2009is,Liao:2010rx,Liao:2010ku,Liao:2010ny,Kanemura:2010bq,Krauss:2011ur}.
\end{enumerate}
Neutrino mass generated from an $n$-loop dimension $d$ diagram 
is estimated as
\begin{align}
m_{\nu} \propto 
\frac{\langle H^{0} \rangle^2}{\Lambda}
\times
\left( \frac{1}{16 \pi^{2}} \right)^{n}
\times 
\epsilon 
\times 
\left(\frac{\langle H^{0} \rangle}{\Lambda}\right)^{d-5},
\end{align}
where $\epsilon$ expresses symbolically the suppression of 
the lepton number violation (suppression mechanism 2). 
Of course, combinations of these mechanisms can be used. For example, in \Ref~\cite{Bonnet:2009ej} one possibility was discussed where neutrino masses are generated by a $d=7$ operator at the two-loop level while small LNV couplings may lead to additional suppression, \ie, all of the above mechanisms are at work. Therefore, studying one possibility systematically also provides important clues on models with combined suppression mechanisms.

In this work, we focus on the complete list of possibilities at one
loop of the neutrino mass operator in \equ{d5} (disregarding
self-energy diagrams), where we consider scalars and fermions as
mediators. That is, we assume neutrino mass suppression mechanism (1),
but we will recover case (2) in some cases. Note that vector-mediated
cases require that the vector should be a gauge boson under a new
symmetry, and the mass should be given by the spontaneous breaking of
the symmetry. This means that the scalar sector of the model needs to
be discussed as well; see \Ref~\cite{Ma:2012xj} for a recent
example. Some of the possibilities found here have been studied
previously in the literature, such as in \Ref~\cite{Ma:1998dn}, but no
complete systematic study has been performed yet.

In \Sec~\ref{sec:decomp}, we list all possible one-loop topologies 
that can lead to neutrino masses, and we discuss the relationship to
the existing literature.  Then in \Sec~\ref{sec:finite}, we study the
conditions for a class of finite contributions to neutrino mass, which
can be interpreted as seesaw extensions. Finally, we
conclude in \Sec~\ref{sec:conclusions}. Detailed lists of the possible
mediator fields for all topologies are provided in
\App~\ref{App:numass}.


\section{Systematic one-loop decomposition}
\label{sec:decomp}

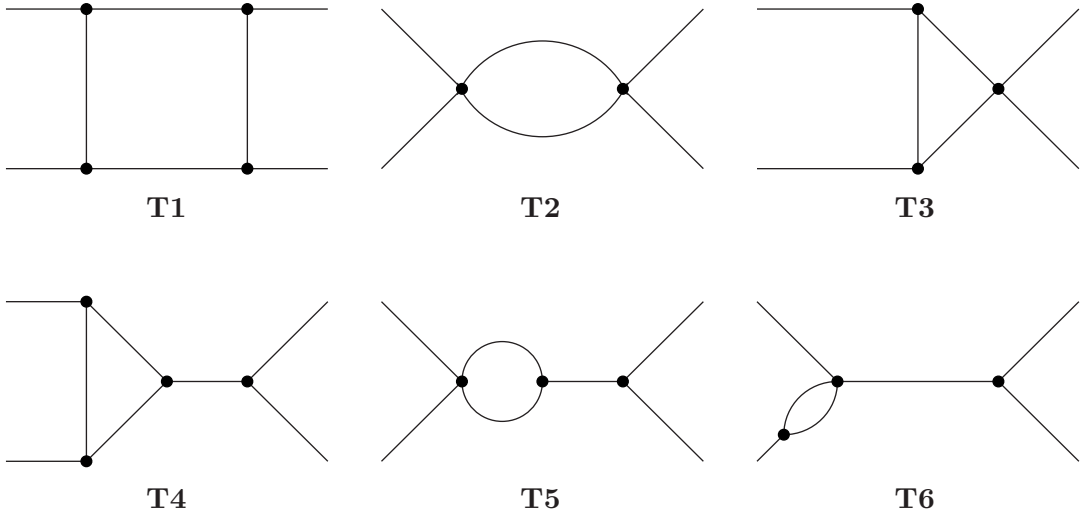
\begin{figure}[t]
\begin{center}
\begin{picture}(420,230)
\Line(10,10)(40,10)
\Line(40,10)(70,40)
\Line(70,40)(40,70)
\Line(40,70)(10,70)
\Line(40,70)(40,10)
\Line(70,40)(100,40)
\Line(100,40)(130,70)
\Line(100,40)(130,10)
\GCirc(40,10){2}{0}
\GCirc(40,70){2}{0}
\GCirc(70,40){2}{0}
\GCirc(100,40){2}{0}
\Text(70,0)[t]{\small{\bf{T4}}}
\Line(150,10)(180,40)
\Line(150,70)(180,40)
\CArc(195,40)(15,0,360)
\Line(210,40)(240,40)
\Line(240,40)(270,10)
\Line(240,40)(270,70)
\GCirc(180,40){2}{0}
\GCirc(210,40){2}{0}
\GCirc(240,40){2}{0}
\Text(210,0)[t]{\small{\bf{T5}}}
\Line(290,10)(300,20)
\Line(290,70)(320,40)
\Line(320,40)(380,40)
\Line(380,40)(410,10)
\Line(380,40)(410,70)
\CArc(320,20)(20,90,180)
\CArc(300,40)(20,270,0)
\GCirc(300,20){2}{0}
\GCirc(320,40){2}{0}
\GCirc(380,40){2}{0}
\Text(350,0)[t]{\small{\bf{T6}}}
\Line(10,120)(130,120)
\Line(10,180)(130,180)
\Line(40,120)(40,180)
\Line(100,120)(100,180)
\GCirc(40,120){2}{0}
\GCirc(40,180){2}{0}
\GCirc(100,120){2}{0}
\GCirc(100,180){2}{0}
\Text(70,110)[t]{\small{\bf{T1}}}
\Line(150,120)(180,150)
\Line(150,180)(180,150)
\Line(240,150)(270,180)
\Line(240,150)(270,120)
\CArc(210,133)(35,25,155)
\CArc(210,167)(35,205,335)
\GCirc(180,150){2}{0}
\GCirc(240,150){2}{0}
\Text(210,110)[t]{\small{\bf{T2}}}
\Line(290,120)(350,120)
\Line(290,180)(350,180)
\Line(350,180)(350,120)
\Line(350,180)(380,150)
\Line(350,120)(380,150)
\Line(380,150)(410,120)
\Line(380,150)(410,180)
\GCirc(350,120){2}{0}
\GCirc(350,180){2}{0}
\GCirc(380,150){2}{0}
\Text(350,110)[t]{\small{\bf{T3}}}
\end{picture}
\caption{\it Topologies of one-loop diagrams with four external legs.}
\label{Fig:one-loopTopo}
\end{center}
\end{figure}

We follow the techniques introduced in \Refs~\cite{Antusch:2008tz,Gavela:2008ra} to find all possible one-loop decompositions of neutrino mass. We therefore start with the key topologies listed in \figu{one-loopTopo}, where we have discarded self-energy-like diagrams. The topologies were found using Feynarts \cite{Hahn:2000kx}. We asked for 4 external legs topologies with no self-energy insertions, containing only 3-legs or 4-legs vertices. At this stage the Lorentz nature (spinor or scalar) of each line was not specified. 
Since we are looking for an operator with two lepton doublets and two Higgs doublets, one can immediately discard topology T2 from dimensional arguments, and therefore we are left with five different topologies.

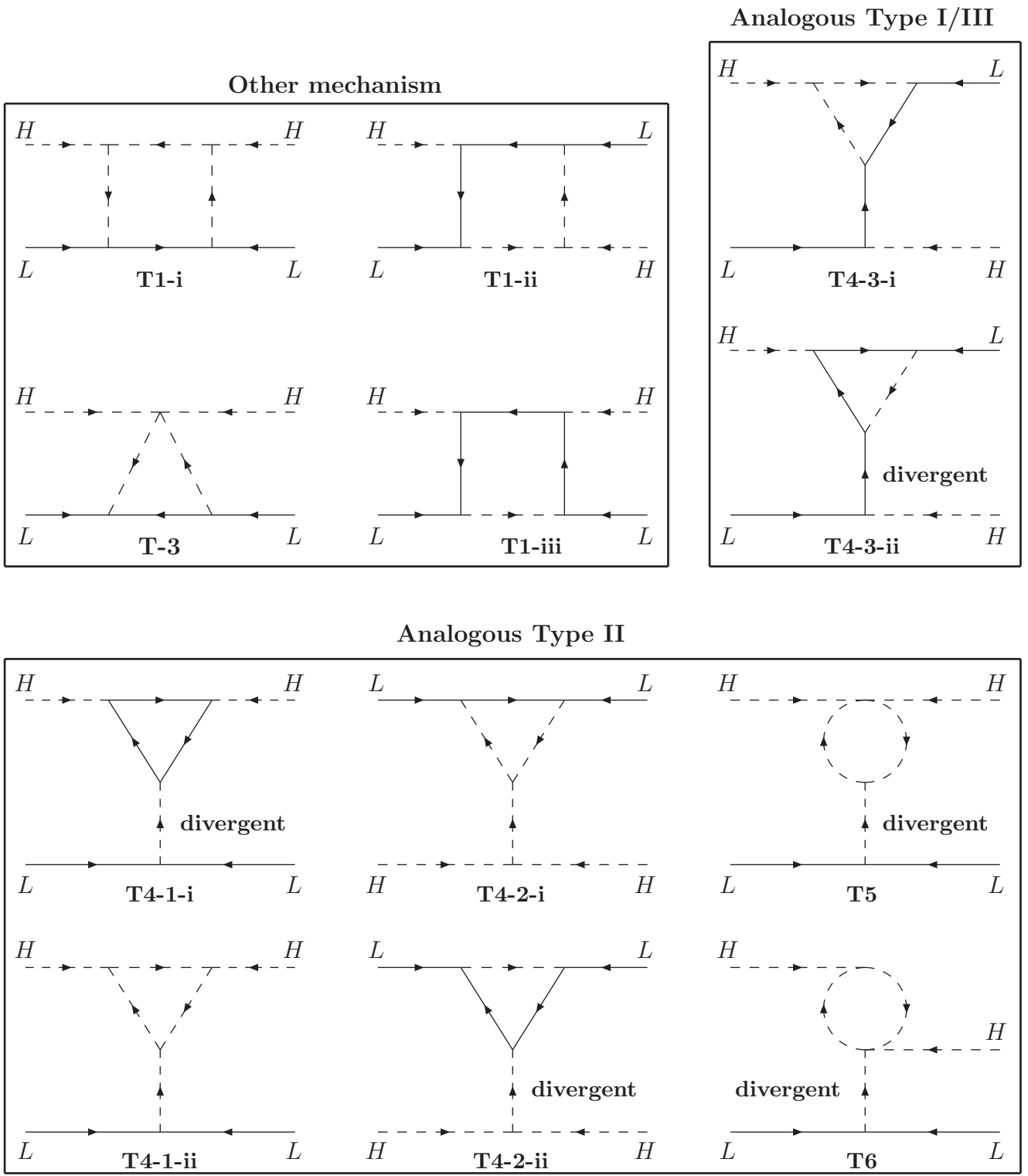
\begin{figure}[pt]
\begin{center}
\begin{picture}(500,500)
\ArrowLine(20,10)(85,10)
\DashArrowLine(20,90)(60,90){5}
\ArrowLine(150,10)(85,10)
\DashArrowLine(150,90)(110,90){5}
\DashArrowLine(60,90)(110,90){5}
\DashArrowLine(85,10)(85,50){5}
\DashArrowLine(85,50)(60,90){5}
\DashArrowLine(110,90)(85,50){5}
\Text(20,5)[t]{$L$}
\Text(150,5)[t]{$L$}
\Text(20,95)[b]{$H$}
\Text(150,95)[b]{$H$}
\Text(85,0)[t]{\small{\bf{T4-1-ii}}}
\DashArrowLine(190,10)(255,10){5}
\ArrowLine(190,90)(230,90)
\DashArrowLine(320,10)(255,10){5}
\ArrowLine(320,90)(280,90)
\DashArrowLine(230,90)(280,90){5}
\DashArrowLine(255,10)(255,50){5}
\ArrowLine(255,50)(230,90)
\ArrowLine(280,90)(255,50)
\Text(190,5)[t]{$H$}
\Text(320,5)[t]{$H$}
\Text(190,95)[b]{$L$}
\Text(320,95)[b]{$L$}
\Text(255,0)[t]{\small{\bf{T4-2-ii}}}
\Text(265,30)[l]{\small{\bf{divergent}}}
\ArrowLine(360,10)(425,10)
\DashArrowLine(360,90)(425,90){5}
\ArrowLine(490,10)(425,10)
\DashArrowLine(490,50)(425,50){5}
\DashArrowLine(425,10)(425,50){5}
\DashArrowArcn(425,70)(20,270,90){5}
\DashArrowArcn(425,70)(20,90,270){5}
\Text(360,5)[t]{$L$}
\Text(490,5)[t]{$L$}
\Text(360,95)[b]{$H$}
\Text(490,55)[b]{$H$}
\Text(425,0)[t]{\small{\bf{T6}}}
\Text(415,30)[r]{\small{\bf{divergent}}}
\ArrowLine(20,140)(85,140)
\DashArrowLine(20,220)(60,220){5}
\ArrowLine(150,140)(85,140)
\DashArrowLine(150,220)(110,220){5}
\ArrowLine(60,220)(110,220)
\DashArrowLine(85,140)(85,180){5}
\ArrowLine(85,180)(60,220)
\ArrowLine(110,220)(85,180)
\Text(20,135)[t]{$L$}
\Text(150,135)[t]{$L$}
\Text(20,225)[b]{$H$}
\Text(150,225)[b]{$H$}
\Text(85,130)[t]{\small{\bf{T4-1-i}}}
\Text(95,160)[l]{\small{\bf{divergent}}}
\DashArrowLine(190,140)(255,140){5}
\ArrowLine(190,220)(230,220)
\DashArrowLine(320,140)(255,140){5}
\ArrowLine(320,220)(280,220)
\ArrowLine(230,220)(280,220)
\DashArrowLine(255,140)(255,180){5}
\DashArrowLine(255,180)(230,220){5}
\DashArrowLine(280,220)(255,180){5}
\Text(190,135)[t]{$H$}
\Text(320,135)[t]{$H$}
\Text(190,225)[b]{$L$}
\Text(320,225)[b]{$L$}
\Text(255,130)[t]{\small{\bf{T4-2-i}}}
\ArrowLine(360,140)(425,140)
\DashArrowLine(360,220)(425,220){5}
\ArrowLine(490,140)(425,140)
\DashArrowLine(490,220)(425,220){5}
\DashArrowLine(425,140)(425,180){5}
\DashArrowArcn(425,200)(20,270,90){5}
\DashArrowArcn(425,200)(20,90,270){5}
\Text(360,135)[t]{$L$}
\Text(490,135)[t]{$L$}
\Text(360,225)[b]{$H$}
\Text(490,225)[b]{$H$}
\Text(425,130)[t]{\small{\bf{T5}}}
\Text(435,160)[l]{\small{\bf{divergent}}}
\ArrowLine(20,310)(60,310)
\DashArrowLine(20,360)(85,360){5}
\ArrowLine(150,310)(110,310)
\DashArrowLine(150,360)(85,360){5}
\ArrowLine(110,310)(60,310)
\DashArrowLine(85,360)(60,310){5}
\DashArrowLine(110,310)(85,360){5}
\Text(20,305)[t]{$L$}
\Text(150,305)[t]{$L$}
\Text(20,365)[b]{$H$}
\Text(150,365)[b]{$H$}
\Text(85,300)[t]{\bf{T-3}}
\ArrowLine(190,310)(230,310)
\DashArrowLine(190,360)(230,360){5}
\ArrowLine(320,310)(280,310)
\DashArrowLine(320,360)(280,360){5}
\DashArrowLine(230,310)(280,310){5}
\ArrowLine(280,360)(230,360)
\ArrowLine(230,360)(230,310)
\ArrowLine(280,310)(280,360)
\Text(190,305)[t]{$L$}
\Text(320,305)[t]{$L$}
\Text(190,365)[b]{$H$}
\Text(320,365)[b]{$H$}
\Text(265,300)[t]{\small{\bf{T1-iii}}}
\ArrowLine(360,310)(425,310)
\DashArrowLine(360,390)(400,390){5}
\DashArrowLine(490,310)(425,310){5}
\ArrowLine(490,390)(450,390)
\ArrowLine(425,310)(425,350)
\ArrowLine(425,350)(400,390)
\ArrowLine(400,390)(450,390)
\DashArrowLine(450,390)(425,350){5}
\Text(360,305)[t]{$L$}
\Text(490,305)[t]{$H$}
\Text(360,395)[b]{$H$}
\Text(490,395)[b]{$L$}
\Text(425,300)[t]{\small{\bf{T4-3-ii}}}
\Text(435,330)[l]{\small{\bf{divergent}}}
\ArrowLine(20,440)(60,440)
\DashArrowLine(20,490)(60,490){5}
\ArrowLine(150,440)(110,440)
\DashArrowLine(150,490)(110,490){5}
\ArrowLine(60,440)(110,440)
\DashArrowLine(110,490)(60,490){5}
\DashArrowLine(60,490)(60,440){5}
\DashArrowLine(110,440)(110,490){5}
\Text(20,435)[t]{$L$}
\Text(150,435)[t]{$L$}
\Text(20,495)[b]{$H$}
\Text(150,495)[b]{$H$}
\Text(85,430)[t]{\small{\bf{T1-i}}}
\ArrowLine(190,440)(230,440)
\DashArrowLine(190,490)(230,490){5}
\DashArrowLine(320,440)(280,440){5}
\ArrowLine(320,490)(280,490)
\DashArrowLine(230,440)(280,440){5}
\ArrowLine(280,490)(230,490)
\ArrowLine(230,490)(230,440)
\DashArrowLine(280,440)(280,490){5}
\Text(190,435)[t]{$L$}
\Text(320,435)[t]{$H$}
\Text(190,495)[b]{$H$}
\Text(320,495)[b]{$L$}
\Text(255,430)[t]{\small{\bf{T1-ii}}}
\ArrowLine(360,440)(425,440)
\DashArrowLine(360,520)(400,520){5}
\DashArrowLine(490,440)(425,440){5}
\ArrowLine(490,520)(450,520)
\DashArrowLine(400,520)(450,520){5}
\ArrowLine(425,440)(425,480)
\DashArrowLine(425,480)(400,520){5}
\ArrowLine(450,520)(425,480)
\Text(360,435)[t]{$L$}
\Text(490,435)[t]{$H$}
\Text(360,525)[b]{$H$}
\Text(490,525)[b]{$L$}
\Text(425,430)[t]{\small{\bf{T4-3-i}}}
\SetWidth{1}
\Line(10,285)(10,510)
\Line(10,285)(330,285)
\Line(10,510)(330,510)
\Line(330,285)(330,510)
\Text(170,517)[b]{\bf{Other mechanism}}
\Line(10,-10)(10,240)
\Line(10,-10)(500,-10)
\Line(10,240)(500,240)
\Line(500,-10)(500,240)
\Text(255,247)[b]{\bf{Analogous Type II}}
\Line(350,285)(350,540)
\Line(350,285)(500,285)
\Line(350,540)(500,540)
\Line(500,285)(500,540)
\Text(425,547)[b]{\bf{Analogous Type I/III}}
\end{picture}
\caption{\it Possible Lorentz structures for topologies T1 to T6. Dashed lines represent scalars, while solid lines represent fermions. Divergent diagrams are marked as such.
}
\label{Fig:Lorentz-Topo}
\end{center}
\end{figure}

For each topology, there are several choices for the assignment of the two lepton doublets and the two Higgs doublets to the four external legs, leading to different types of fields running in the loop: scalars or fermions. The different possibilities are listed in \figu{Lorentz-Topo}. The corresponding neutrino masses obtained after EWSB and the possible different hypercharge assignments are listed in Appendix~\ref{App:numass}. Note that we classify the new fields by their SM gauge charges using $\bf{X}^{\mathcal{L}}_Y$, where
\begin{itemize}
\item $\bf{X}$ denotes the $SU(2)$ nature, {\it i.e.}, singlet $\bf 1$, doublet $\bf 2$ or triplet $\bf 3$,
\item $\mathcal{L}$ refers to the Lorentz nature,  {\it i.e.}, scalar ($S$) or fermion ($F$),
\item $Y \equiv 2(Q - I_3)$ refers to the hypercharge. 
\end{itemize}
Since the lepton doublets and Higgs doublets are neutral
under $SU(3)_C$, we don't introduce here any color indices and
charges. However, each new field introduced can have a non-zero color
charge. Since the outside of all diagrams contains only Higgs and 
lepton fields, which are color neutral, color indices on the internal lines 
can only appear in combinations such as ${\bf 3}+{\bf \bar 3}$, 
${\bf 6}+{\bf \bar 6}$,  ${\bf 8}+{\bf 8}$ etc. Thus, as soon as one 
fixes the color index of one of the internal fields, all other fields 
are automatically (and trivially) fixed. We therefore do not list all 
possible combinations of color
(For models with colored mediator, see {\it e.g.}, Ref.~\cite{FileviezPerez:2009ud,Choubey:2012ux}). 
In terms of the above defined notation, the
mediators in \figu{TreeSeesaw} can be written as $1^F_0$, $3^S_{-2}$,
and $3^F_0$, whereas the SM Higgs and lepton doublets can be written
as $2^S_{+1}$ and $2^F_{-1}$, respectively. On the other hand, not
every $2^S_{+1}$ can be interpreted as a Higgs doublet, since the
participation in EWSB depends on the scalar potential, which we do not
study in this work, \ie, having the right SM quantum numbers is just a
necessary condition.

The results from \figu{Lorentz-Topo} and the tables in the Appendix
can be summarized as follows. T1 and T4 can be realized with different
Lorentz structures, while the Lorentz nature of topologies T3, T5, and
T6 is unique. Topologies T1 and T3
are irreducible (box ``other mechanisms'' in the figure), \ie, they cannot be separated by cutting one
propagator. In particular, diagrams T1-ii (Zee
model~\cite{Zee:1980ai}\footnote{%
Zee model includes two Higgs doublets.}), 
T1-iii, and T3 have been explicitely
discussed in \Ref~\cite{Ma:1998dn}. Diagram T1-i has been discussed in
the context of the supersymmetric version of the dark doublet
model~\cite{Kubo:2006yx,Ma:2006km}. Since these have been extensively studied 
in the literature
(although not all possible versions, see the Appendix), we will not
discuss them in greater detail. However, we note that 
mediators that are also present in one of the tree level
seesaws in Fig.~\ref{Fig:TreeSeesaw} appear in some cases, see
Tables~\ref{tab:massandchargeT1} and \ref{tab:massandchargeT3T5T6} in
the Appendix for details. In these cases, it is easy to show that
charging the matter fields under a simple $\mathbb{Z}_2$ symmetry is
enough to forbid the couplings leading to tree level neutrino
masses. Therefore, for each set of mediators, the one loop
contribution can be made the leading contribution to neutrino mass
over the tree level one.

\begin{figure}[t]
\begin{center}
\begin{picture}(500,120)
\ArrowLine(20,10)(85,10)
\ArrowLine(85,10)(85,90)
\ArrowLine(150,90)(85,90)
\DashArrowLine(20,90)(85,90){5}
\DashArrowLine(150,10)(85,10){5}
\Text(20,5)[t]{$L$}
\Text(150,5)[t]{$H$}
\Text(20,95)[b]{$H$}
\Text(150,95)[b]{$L$}
\Text(95,50)[l]{$1^F_0 / 3^F_0$}
\Text(85,0)[t]{\small{$Y_{\nu}$}}
\Text(85,105)[b]{\small$Y_{\nu}^{\rm{loop}}$}
\SetWidth{2}
\GCirc(85,90){10}{0.6}
\SetWidth{0.5}
\ArrowLine(190,10)(255,10)
\DashArrowLine(255,10)(255,90){5}
\DashArrowLine(320,90)(255,90){5}
\DashArrowLine(190,90)(255,90){5}
\ArrowLine(320,10)(255,10){5}
\Text(190,5)[t]{$L$}
\Text(320,5)[t]{$L$}
\Text(190,95)[b]{$H$}
\Text(320,95)[b]{$H$}
\Text(265,50)[l]{$3^S_{-2}$}
\Text(255,0)[t]{\small{$Y_{\Delta}$}}
\Text(255,105)[b]{\small$\mu_{\Delta}^{\rm{loop}}$}
\SetWidth{2}
\GCirc(250,90){10}{0.6}
\SetWidth{0.5}
\ArrowLine(360,10)(425,10)
\DashArrowLine(425,10)(425,90){5}
\DashArrowLine(490,90)(425,90){5}
\DashArrowLine(360,90)(425,90){5}
\ArrowLine(490,10)(425,10){5}
\Text(360,5)[t]{$L$}
\Text(490,5)[t]{$L$}
\Text(360,95)[b]{$H$}
\Text(490,95)[b]{$H$}
\Text(435,50)[l]{$3^S_{-2}$}
\Text(425,0)[t]{\small{$Y_{\Delta}^{\rm{loop}}$}}
\Text(425,105)[b]{\small$\mu_{\Delta}$}
\SetWidth{2}
\GCirc(420,10){10}{0.6}
\end{picture}
\caption{\it Type I/III (left) and type II (center and right) seesaw realizations via one loop vertices.
}
\label{Fig:LoopSeesaw}
\end{center}
\end{figure}
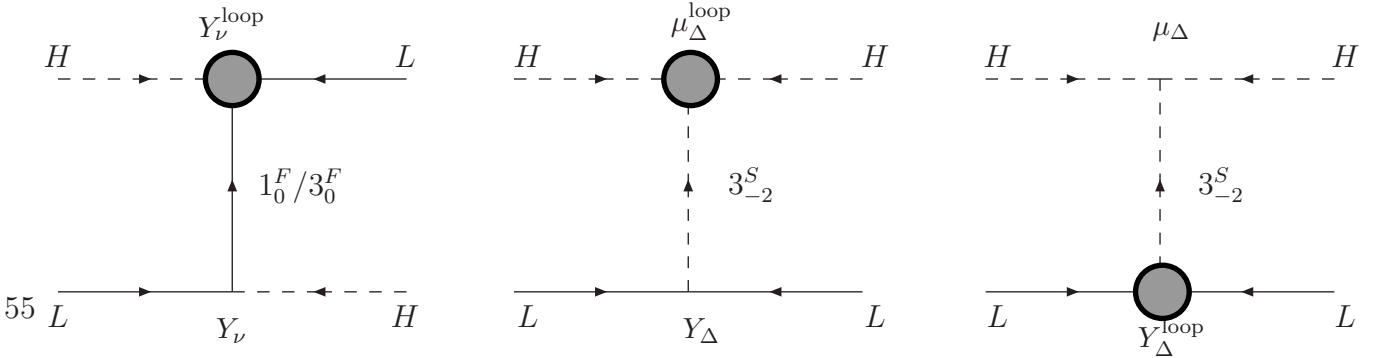

Topologies T4, T5, and T6 can be interpreted as extensions of the usual seesaw mechanisms at one loop, where we mark the extensions of the type~I/III and type~II seesaws by the boxes in  \figu{Lorentz-Topo}. These extensions are illustrated in \figu{LoopSeesaw}, where one of the vertices is generated at one loop. For the type~I/III seesaw (left panel), there is only one type of vertex and therefore only one possibility to modify these seesaws. For the type~II seesaw, the vertex with the external scalars (middle panel) or fermions (right panel) can be modified. In fact, in T4-1, T5, and T6, the vertex with the external scalars in generated at one loop, whereas in T4-2 the vertex with the fermions is generated at one loop.

It turns out that the corresponding T4, T5, and T6 diagrams can be divided into two categories: the diagrams that lead to a finite contribution to neutrino masses and the one that present a divergence, where we mark the divergent ones in \figu{Lorentz-Topo}. For T4, we list the finite ones (T4-1-ii, T4-2-i, T4-3-i) in Table~\ref{tab:massandchargeT4fin} and the divergent ones (T4-1-i, T4-2-ii, T4-3-ii) in Table~\ref{tab:massandchargeT4div}. T5 and T6 turn out to be divergent, as shown in the Appendix, see Table~\ref{tab:massandchargeT3T5T6}.
We will discuss the finite possibilities as seesaw extensions at one loop in \Sec~\ref{sec:finite}. 


But what about the divergent ones? There have to be  counter-terms to absorb such divergences in order to obtain finite neutrino masses. In all these cases, the counter-terms are actually the tree-level realizations of the $d=5$ operator.  In order to illustrate that, consider the diagram T4-1-i.
We introduce the following Lagrangian, 
which is written with component fields:
\begin{eqnarray}
\mathscr{L}_{\text{{\sf T4}-1-i}}
=&
\left[
y_{L} (\overline{\psi'} {\rm P}_{L} \psi \Delta^{0})
+
Y_{L} (\overline{\psi} {\rm P}_{L} \Psi H^{0})
+
Y'_{R} (\overline{\Psi} {\rm P}_{R} \psi' H^{0})
+
Y_{\Delta} \overline{\nu^C_L}\nu_L\Delta^{0*}
+
{\rm H.c.}
\right]
\nonumber 
\\
&
+
M_{\psi} \overline{\psi} \psi
+
M'_{\psi'} \overline{\psi'} \psi'
+
\mathcal{M}_{\Psi} \overline{\Psi} \Psi
+
m_{\Delta}^{2} \Delta^{0*} \Delta^{0}.
\end{eqnarray}
This Lagrangian can be understood as the components of the following
SM gauge invariant realization,
\begin{align}
\mathscr{L}_{\text{{\sf T4}-1-i}}
\subset&
\left[
y_{L} 
(\overline{\psi'}^{i} {(\tau^{a})_{i}}^{j} {\rm P}_{L} \psi_{j} \Delta^{a})
+
Y_{L} (\overline{\psi}^{i} {\rm P}_{L} \Psi H_{i})
+
Y'_{R} (\overline{\Psi} {\rm P}_{R} \psi'_{i} ({\rm i} \tau^{2})^{ij} H_{j})
+
Y_{\Delta} \overline{L^C}_{i} 
({\rm i} \tau^{2} \tau^{a})^{ij} L_{j} \Delta^{\dagger a}
+
{\rm H.c.}
\right]
\nonumber 
\\
&+
M_{\psi} \overline{\psi}^{i} \psi_{i}
+
M'_{\psi'} \overline{\psi'}^{i} \psi'_{i}
+
\mathcal{M}_{\Psi} \overline{\Psi} \Psi
+
m_{\Delta}^{2} \Delta^{\dagger a} \Delta^{a},
\label{eq:LT41i-realization}
\end{align}
where $\psi$ and $\psi'$ are given as $SU(2)_{L}$ doublet fields,
and $\Psi$ is a singlet field.
The scalar field $\Delta^{a}$ has the same charges of the SM gauge
symmetries as the triplet scalar field that appears in the type II 
seesaw model.
The indices $i$ and $j$ indicate $SU(2)_{L}$ fundamental
representation ($i,j=1,2$), 
and $a$ is for the adjoint representation ($a=1,2,3$).
There are other realizations than that shown in 
Eq.~\eqref{eq:LT41i-realization}. 
All the possibilities are listed in Table \ref{tab:massandchargeT4div}.

As shown in Table~\ref{tab:massandchargeT4div}, one finds that the neutrino mass is divergent. We performed the computation in dimensional regularization therefore the $1/\epsilon$ pole corresponds to a logarithmic divergence in the momentum integral. 
 This divergence can however be absorbed by the tree level counter-terms coming from the regular tree level type~II seesaw. 
Together with the couplings of the leptons doublets to the scalar triplet $\Delta$ (see Fig.~\ref{Fig:TreeSeesaw}) they lead to 
\begin{eqnarray}
m_{\nu_{\rm tree}}=-4Y_{\Delta}\mu_{\Delta}\frac{\langle H^0\rangle^2}{M_{\Delta}^2},
\end{eqnarray}
in a classical type~II Seesaw. In our case, we can use this tree level counter term to cancel the divergences arising  in the one-loop computation. Using a usual renormalization procedure, imposing that the denominator of the neutrino propagator be zero at the physical mass $m_{\nu_{\rm phys}}$
\begin{equation}
\left.\left[p^2-(m_{\nu_{\rm tree}}+m_{\nu_{\rm loop}})^2\right]\right|_{p^2=m^2_{\nu_{\rm phys}}} =0,
\end{equation}
and writing $m_{\nu_{\rm tree}}=m_{\nu_{\rm phys}}+\delta m_{\nu_{\rm CT}}$, we obtain $ \delta m_{\nu_{\rm CT}}= \frac{1}{2}m_{\nu_{\rm loop}}$.

It is easy to see that the renormalization of the neutrino mass
corresponds actually to the vertex renormalization of
$\mu_{\Delta}$. Therefore, the divergent diagrams do not present new
possibilities to generate neutrino mass per se, and can be regarded as
corrections to the tree level seesaws. This is consistent with the
fact that one cannot forbid the tree level
contribution in these cases with just a discrete symmetry, as we will demonstrate for the finite
cases in the next section. \footnote{We note in passing that a neutrino 
mass model along these lines of argument, but based on diagram 
T4-1-ii, has recently been proposed in \cite{Kanemura:2012rj}.}


\section{Loop extensions of the seesaws}
\label{sec:finite}


\subsection{Conditions for a genuine one-loop seesaw}

Here we discuss the class of diagrams yielding finite contributions to neutrino mass. They can be regarded as extensions of the tree level seesaws in \figu{TreeSeesaw}, where one vertex in the tree level diagram is generated via a loop. Diagrams T4-1-ii and T4-2-i are linked to the type~II seesaw, while diagram T4-3-i is linked to the type~I and type~III seesaws, as indicated by the boxes in \figu{Lorentz-Topo}. 

The most important question is: if we want to have the loop diagram as leading contribution to neutrino mass, can we genuinely forbid the tree level contribution to the $d=5$ operator, such as by a $U(1)$ or discrete symmetry? As we discussed earlier, this is always possible for topologies T1 and T3. To be more explicit, the conditions for the matter field charges to forbid the tree level diagram can be read off from \figu{TreeSeesaw}:
\begin{eqnarray}
\text{Type~I~seesaw: }&q_L+q_H-q_N \neq 0 \, ,\label{CondTypeI}\\
\text{Type~II~seesaw: }&2q_L-q_{\Delta} \neq 0 \quad \mathrm{or} \quad 2 q_H + q_\Delta \neq 0 \, ,\label{CondTypeII}\\
\text{Type~III~seesaw: }&q_L+q_H-q_\Sigma \neq 0,\label{CondTypeIII}
\end{eqnarray}
where $q_{X}$ is a charge of field $X$ under the new symmetry.
Using notation of Table~\ref{tab:massandchargeT4fin} (in the Appendix) we can write the conditions for the interactions  under a new extra symmetry for the finite diagrams of T4 as:
\begin{eqnarray}
{\small \text{\bf T4-1-ii}}: \begin{array}{ccc}
2q_L-q_\Delta=0 &&\\
\left.\begin{array}{c}
q_\Delta-q_\phi+q_{\phi'}=0\\
q_H+q_\Phi-q_{\phi'}=0\\
q_H+q_\phi-q_{\Phi}=0
\end{array}
\right\}&\Rightarrow & 2q_H+q_\Delta=0
\end{array}
\end{eqnarray}
\begin{eqnarray}
{\small \text{\bf T4-2-i}}: \begin{array}{ccc}
2q_H+q_\Delta=0 &&\\
\left.\begin{array}{c}
-q_\Delta-q_\phi+q_{\phi'}=0\\
q_L+q_\psi-q_{\phi'}=0\\
q_L+q_\phi-q_{\psi}=0
\end{array}
\right\}&\Rightarrow & 2q_L-q_\Delta=0
\end{array}
\end{eqnarray}
\begin{eqnarray}
{\small \text{\bf T4-3-i}} : \begin{array}{ccc}
q_L+q_H-q_{\Psi}=0 &&\quad (\Psi\sim1^S_0\,\rm{or}\,3^S_0)\\
\left.\begin{array}{c}
q_{\Psi}+q_\psi-q_{\phi}=0\\
q_H+q_\phi-q_{\phi'}=0\\
q_L+q_{\phi'}-q_{\psi}=0
\end{array}
\right\}&\Rightarrow & q_L+q_H-q_{\Psi}=0
\end{array}
\end{eqnarray}
It is clear that these conditions are in contradictions with Eqs.~(\ref{CondTypeI}), (\ref{CondTypeII}), and (\ref{CondTypeIII}). As a consequence, the finite loop diagrams of T4 will come with a  tree level contribution, which will be generically leading.

For example, consider the case T4-3-i. Comparing the contribution from this diagram to the type~I ($\Psi=N$) or type~III ($\Psi=\Sigma$) tree level seesaw, one has to compare (\cf, Fig.~\ref{Fig:TreeSeesaw})
\begin{eqnarray}
Y_{\nu}\frac{1}{M_\Psi}Y_{\nu}^T\,
\end{eqnarray}
to (\cf, Table~\ref{tab:massandchargeT4fin})
\begin{eqnarray}
\mnuTfourthreei\,,
\end{eqnarray}
which is tantamount as comparing $Y_{\nu}$ to $y y' M_{\psi} I_3$.
As a matter of fact all the loop contributions considered in this part
are proportional to the function $I_3$ times a product of several
couplings (see Appendix~\ref{App:numass}). A quick evaluation of this
function shows that if the masses of the new fields are above
$100~\rm{GeV}$, $|I_3|\lesssim0.1/(\rm{TeV}^2)$ and becomes rapidly very
small:
$|I_3(\sim1~\rm{TeV},\sim1~\rm{TeV},\sim1~\rm{TeV})|\lesssim0.001/(\rm{TeV}^2)$. As
a consequence, except from unnatural suppression of the couplings
involved in the corresponding tree diagram, the tree-level contribution 
will always be dominant.

\begin{figure}[t]
\begin{center}
\begin{picture}(370,110)
\DashArrowLine(20,10)(85,10){5}
\ArrowLine(20,90)(60,90)
\DashArrowLine(150,10)(85,10){5}
\ArrowLine(150,90)(110,90)
\ArrowLine(60,90)(85,90)
\ArrowLine(110,90)(85,90)
\DashArrowLine(85,10)(85,50){5}
\DashArrowLine(85,50)(60,90){5}
\DashArrowLine(110,90)(85,50){5}
\Text(70,95)[b]{$\psi$}
\Text(100,95)[b]{$\psi^c$}
\Text(58,70)[r]{$\phi\sim2^S_1$}
\Text(105,70)[l]{$\phi^{\dagger}$}
\Text(20,5)[t]{$H$}
\Text(150,5)[t]{$H$}
\Text(20,95)[b]{$L$}
\Text(150,95)[b]{$L$}
%
\Boxc(85,80)(170,50)
\ArrowLine(220,10)(260,10)
\ArrowLine(350,10)(310,10)
\ArrowLine(260,10)(285,10)
\ArrowLine(310,10)(285,10)
\DashLine(260,10)(260,50){5}
\DashLine(310,10)(310,50){5}
\Line(255,45)(265,55)
\Line(255,55)(265,45)
\Line(315,45)(305,55)
\Line(315,55)(305,45)
\Text(220,15)[b]{$L$}
\Text(350,15)[b]{$L$}
\Text(270,15)[b]{$\psi$}
\Text(300,15)[b]{$\psi^c$}
\Text(260,60)[b]{$\left<\phi^0\right>$}
\Text(310,60)[b]{$\left<\phi^0\right>$}
\LongArrow(175,80)(210,30)
\end{picture}
\end{center}
\caption{\it Illustration of T4-2-i with a Majorana fermion inside the loop, leading to a type~I or type~III seesaw if the scalar doublet can obtain a vev.}
\label{Fig:LN1}
\end{figure}
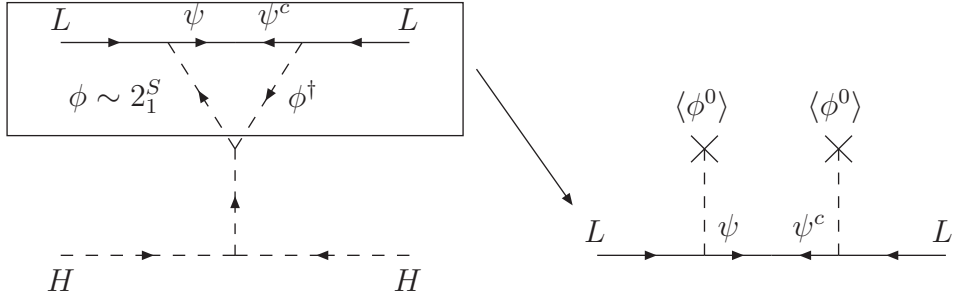

However, there is a way to forbid the tree level diagrams, which was
adopted in the dark doublet model~\cite{Ma:2006km}: one can promote the fermions inside the loop to be Majorana fermions and assume that all couplings are lepton number conserving.  This way one allows the lepton number to be broken only by Majorana masses and not by any couplings.
For example, one can generate T4-2-i by imposing the fermion $\psi$ to run in the loop to be a Majorana fermion, see \figu{LN1} for an illustration. If one assumes that all couplings conserve lepton number, the tree level coupling $\Delta LL$ does not  exist and the loop contribution becomes the leading order (which does not contain lepton number violating couplings). This option reduces the number of fields and hypercharge assignment to only two possibilities: $\psi \sim 1^F_0$ or $3^F_0$.
However, as can be seen from \figu{LN1} this leads automatically to the tree level realization of the type~I or type~III seesaw if the new scalar $\phi$ develops a vev. Imposing a $\mathbb{Z}_2$ symmetry $q_{\psi}=q_{\phi}=1$, $q_{\rm{others}}=0$ prevents this, and makes the loop diagram the leading order for neutrino mass. Note that this case can be interpreted as generalized type~II seesaw, where the lepton number violating coupling is generated by a loop diagram.

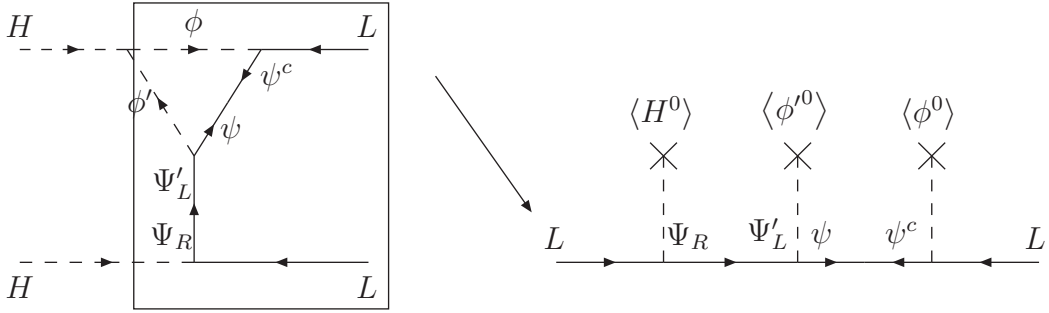
\begin{figure}[t]
\begin{center}
\begin{picture}(420,110)
\DashArrowLine(20,10)(85,10){5}
\DashArrowLine(20,90)(60,90){5}
\ArrowLine(150,10)(85,10)
\ArrowLine(150,90)(110,90)
\DashArrowLine(60,90)(110,90){5}
\ArrowLine(85,10)(85,50)
\DashArrowLine(85,50)(60,90){5}
\ArrowLine(110,90)(97.5,70)
\ArrowLine(85,50)(97.5,70)
\Text(85,95)[b]{$\phi$}
\Text(70,70)[r]{$\phi'$}
\Text(110,80)[l]{$\psi^c$}
\Text(95,60)[l]{$\psi$}
\Text(85,20)[r]{$\Psi_R$}
\Text(85,40)[r]{$\Psi'_L$}
\Text(20,5)[t]{$H$}
\Text(150,5)[t]{$L$}
\Text(20,95)[b]{$H$}
\Text(150,95)[b]{$L$}
%
\Boxc(110,50)(95,115)
\ArrowLine(220,10)(260,10)
\ArrowLine(400,10)(360,10)
\ArrowLine(260,10)(310,10)
\ArrowLine(310,10)(335,10)
\ArrowLine(360,10)(335,10)
\DashLine(260,10)(260,50){5}
\DashLine(310,10)(310,50){5}
\DashLine(360,10)(360,50){5}
\Line(255,45)(265,55)
\Line(255,55)(265,45)
\Line(315,45)(305,55)
\Line(315,55)(305,45)
\Line(355,45)(365,55)
\Line(355,55)(365,45)
\Text(220,15)[b]{$L$}
\Text(400,15)[b]{$L$}
\Text(270,15)[b]{$\Psi_R$}
\Text(300,15)[b]{$\Psi'_L$}
\Text(320,15)[b]{$\psi$}
\Text(350,15)[b]{$\psi^c$}
\Text(260,60)[b]{$\left<H^0\right>$}
\Text(310,60)[b]{$\left<{\phi'}^0\right>$}
\Text(360,60)[b]{$\left<{\phi}^0\right>$}
\LongArrow(175,80)(210,30)
\end{picture}
\end{center}
\caption{\it 
 Illustration of T4-3-i with a Majorana fermion inside the loop, leading to a $d=7$ operator if the scalars can obtain a vev.}
\label{Fig:LN2}
\end{figure}

A similar approach can be used for diagram T4-3-i, see \figu{LN2}. Here
$\psi$ within the loop is promoted to a Majorana fermion when possible
(here there is more than one possibility, see
Table~\ref{tab:massandchargeT4fin}), whereas  $\Psi$ is to be a Dirac
4-spinor $\Psi~\left(\Psi'_L,\Psi_R\right)$. In this case, the couplings
are assumed to be lepton number conserving as well, and the tree level
type~I or type~III seesaw contribution is forbidden. Then one needs
again an extra $\mathbb{Z}_2$ symmetry to prevent $\phi$ and $\phi'$
from getting a vev to avoid any other tree level diagram, see
Fig.~\ref{Fig:LN2}.

Finally, for the case T4-1-ii, only scalars are present in the loop, which means that the above mechanism cannot be used to prevent the tree level contribution.

As a conclusion, requiring  Majorana fermions in the loop (only few possibilities of the one listed in Appendix \ref{App:numass}), lepton number conserving couplings, and an extra discrete symmetry to forbid extra scalars to develop a vev are the essential ingredients to have a genuine one loop seesaw mechanism for topology~4.


\subsection{Relationship to linear and inverse seesaw}

Although the tree-level seesaw allows one to generate small neutrino masses, its tree-level realization suffers from not having sizable signatures at low-energy (via lepton flavour violation (LFV) processes) or high-energy (via the production of the seesaw mediators at the LHC). If the neutrino mass in the type~I or type~III seesaw is 
\begin{equation}
m_{\nu}=- Y^T_{\nu}\frac{\langle H^0 \rangle^2}{M}Y_{\nu}\,,
\label{eq:TypeI_III_mass}
\end{equation}
where $M$ is the mass of the fermionic singlet (type I) or triplet (type III), the LFV processes generated by the presence of these mediators are driven by the quantity~\cite{Abada:2007ux}
\begin{equation}
Y^{\dagger}_{\nu}\frac{\langle H^0 \rangle^2}{M^2}Y_{\nu}\,.
\label{eq:TypeI_III_LFV}
\end{equation}
Since neutrino masses are constrained to be very small, Eq.~(\ref{eq:TypeI_III_mass}) implies that, barring unusual cancellations, that either $M$ is very high or $Y_{\nu}$ is very small. If $M$ is very high, the mediator cannot be produced at the LHC. If $Y_{\nu}$ is very small, the mediator is not detectable at the LHC, and LFV processes are expected to be small as well.

One way to escape this is to use the fact that neutrino masses violate lepton number,  while LFV processes do not (see Ref.~\cite{Abada:2007ux} for a review in the effective operator framework). This new mechanism is often called inverse seesaw~\cite{Mohapatra:1986bd} or linear seesaw~\cite{Akhmedov:1995ip,Akhmedov:1995vm}.
In the inverse seesaw, two singlet fermions\footnote{Note that one can do the same for the type~III seesaw by having to different triplets.} (2-spinors)
$N_1$ and $N_2$ 
are proposed. If lepton number is conserved, one has
\begin{eqnarray}
\begin{array}{cc}
 & \begin{array}{ccc} \nu &\quad\quad\quad N_1^{C} \quad\quad\quad & N_2 \end{array}\\
 &\\
  \begin{array}{c} \overline{\nu^C} \\ \overline{N_1} \\ \overline{N_2^{C}} \end{array} &
  \left(\begin{array}{ccc}
  0 & Y_{\nu} \langle H^0\rangle & 0\\
  Y^T_{\nu} \langle H^0\rangle & 0 & M\\
  0 & M^{T}&0
  \end{array}\right) \, ,
\end{array}
\end{eqnarray}
which obviously does not lead to neutrino masses.  Then, one can turn
on either a small Majorana mass $\mu$ for $N_2$ (inverse seesaw) or a 
small lepton number violating coupling $\epsilon$ (linear 
seesaw), or both, 
\begin{eqnarray}
\begin{array}{cc}
 & \begin{array}{ccc} \nu &\quad\quad\quad N_1^{C} \quad\quad\quad & N_2 \end{array}\\
 &\\
  \begin{array}{c} \overline{\nu^C} \\ \overline{N_1} \\ \overline{N_2^C} \end{array} &
  \left(\begin{array}{ccc}
  0 & Y_{\nu} \langle H^0\rangle & \epsilon Y'_{\nu}\langle H^0\rangle\\
  Y^T_{\nu} \langle H^0\rangle & 0 & M\\
  \epsilon Y_{\nu}'^{T} \langle H^{0} \rangle & M^{T} &  \mu
  \end{array}\right) \, ,
\end{array}
\end{eqnarray}
to generate a small Majorana mass for the light neutrinos:
\begin{equation}
m_{\nu}=\epsilon\left( {Y'_{\nu}}^T\frac{\langle H^0 \rangle^2}{M}Y_{\nu}+Y_{\nu}^T\frac{\langle H^0 \rangle^2}{M}Y'_{\nu}\right)-Y^T_{\nu}\frac{\mu \langle H^0 \rangle^2}{M^2}Y_{\nu}\,.
\label{equ:InvSS_mass}
\end{equation}
Since lepton number is accidentally conserved in the Standard Model, it may be natural that any lepton number violating parameter is small because it is protected by the symmetry. As a consequence, the masses of the mediators might at the TeV scale, while the Yukawa couplings should be sufficiently large. Since the combination driving the LFV processes remains the same as in Eq.~(\ref{eq:TypeI_III_LFV}), it  can be also large.
This mechanism is also naturally implemented in the type~II seesaw, where the coupling $\mu_{\Delta}$ between triplet scalar and Higgses usually plays the role of the lepton number violation in order to have large LFV processes, driven by the other coupling $Y^{\dagger}_{\Delta}\frac{1}{M^2}Y_{\Delta}$. Therefore, the mass of the heavy fields can still be at the TeV scale.

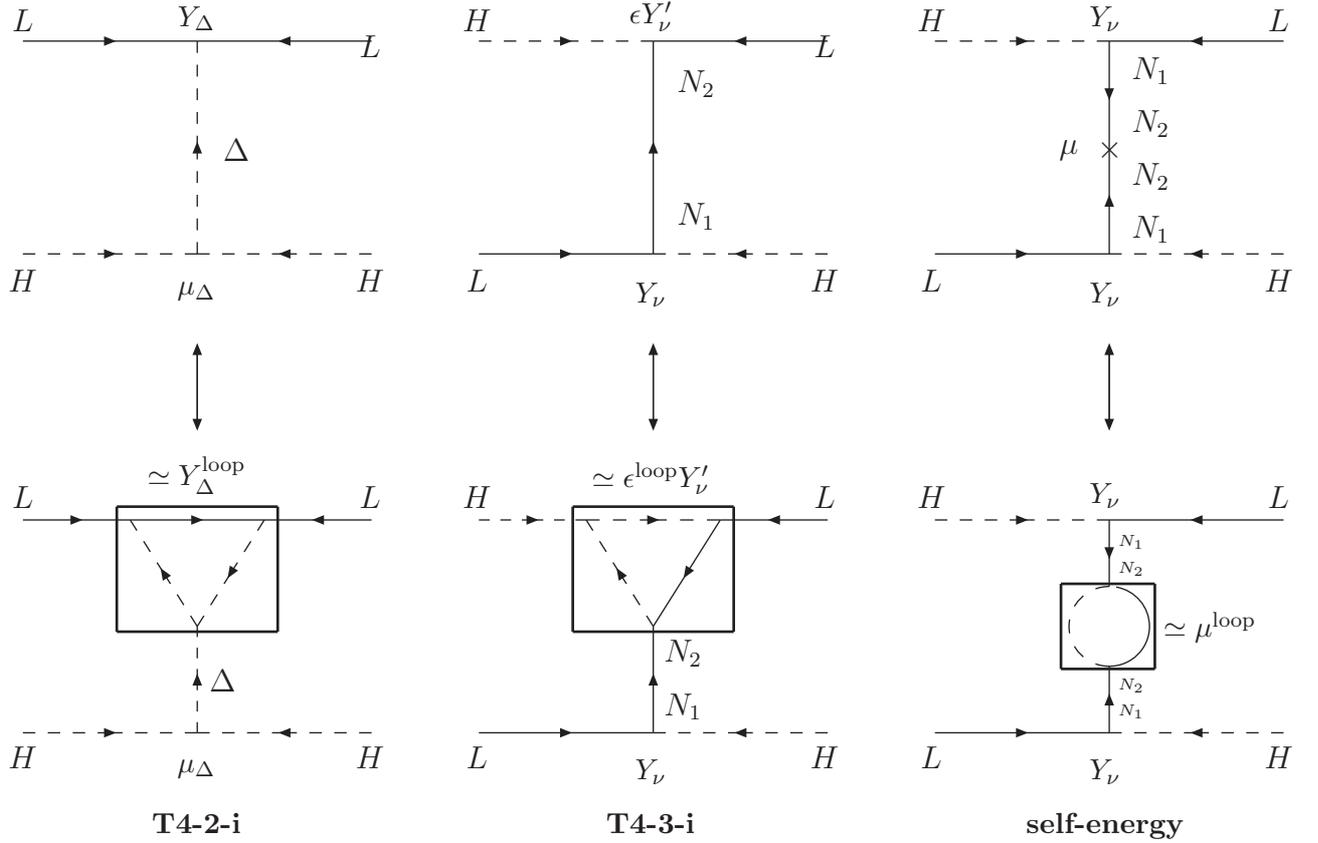
\begin{figure}[t]
\begin{center}
\begin{picture}(500,330)
\DashArrowLine(20,40)(85,40){5}
\ArrowLine(20,120)(60,120)
\DashArrowLine(150,40)(85,40){5}
\ArrowLine(150,120)(110,120)
\ArrowLine(60,120)(110,120)
\DashArrowLine(85,40)(85,80){5}
\DashArrowLine(85,80)(60,120){5}
\DashArrowLine(110,120)(85,80){5}
\Text(20,35)[t]{$H$}
\Text(150,35)[t]{$H$}
\Text(20,125)[b]{$L$}
\Text(150,125)[b]{$L$}
\Text(90,60)[l]{$\Delta$}
\Text(85,30)[t]{\small{$\mu_{\Delta}$}}
\Text(85,130)[b]{\small{$\simeq Y_{\Delta}^{\rm{loop}}$}}
\Text(85,10)[t]{\small{\bf{T4-2-i}}}
\SetWidth{1}
\Line(55,78)(115,78)
\Line(55,125)(115,125)
\Line(55,78)(55,125)
\Line(115,78)(115,125)
\SetWidth{0.5}
\ArrowLine(190,40)(255,40)
\DashArrowLine(190,120)(230,120){5}
\DashArrowLine(320,40)(255,40){5}
\ArrowLine(320,120)(280,120)
\DashArrowLine(230,120)(280,120){5}
\ArrowLine(255,40)(255,80)
\DashArrowLine(255,80)(230,120){5}
\ArrowLine(280,120)(255,80)
\Text(190,35)[t]{$L$}
\Text(320,35)[t]{$H$}
\Text(190,125)[b]{$H$}
\Text(320,125)[b]{$L$}
\Text(260,50)[l]{$N_1$}
\Text(260,70)[l]{$N_2$}
\Text(255,30)[t]{\small{$Y_{\nu}$}}
\Text(255,130)[b]{\small{$\simeq\epsilon^{\rm{loop}}Y'_{\nu}$}}
\Text(255,10)[t]{\small{\bf{T4-3-i}}}
\SetWidth{1}
\Line(225,78)(285,78)
\Line(225,125)(285,125)
\Line(225,78)(225,125)
\Line(285,78)(285,125)
\SetWidth{0.5}
\ArrowLine(360,40)(425,40)
\ArrowLine(425,40)(425,65)
\ArrowLine(425,120)(425,95)
\CArc(425,80)(15,270,80)
\DashCArc(425,80)(15,90,270){5}
\ArrowLine(490,120)(425,120)
\DashArrowLine(360,120)(425,120){5}
\DashArrowLine(490,40)(425,40){5}
\Text(360,35)[t]{$L$}
\Text(490,35)[t]{$H$}
\Text(360,125)[b]{$H$}
\Text(490,125)[b]{$L$}
\Text(430,112)[l]{\tiny{$N_1$}}
\Text(430,48)[l]{\tiny{$N_1$}}
\Text(430,102)[l]{\tiny{$N_2$}}
\Text(430,58)[l]{\tiny{$N_2$}}
\Text(425,30)[t]{\small{$Y_{\nu}$}}
\Text(425,125)[b]{\small$Y_{\nu}$}
\Text(447,80)[l]{\small$\simeq \mu^{\rm{loop}}$}
\Text(425,10)[t]{\small{\bf{self-energy}}}
\SetWidth{1}
\Line(407,64)(442,64)
\Line(407,96)(442,96)
\Line(407,64)(407,96)
\Line(442,64)(442,96)
\SetWidth{0.5}
\DashArrowLine(20,220)(85,220){5}
\DashArrowLine(85,220)(85,300){5}
\ArrowLine(150,300)(85,300)
\ArrowLine(20,300)(85,300)
\DashArrowLine(150,220)(85,220){5}
\Text(20,215)[t]{$H$}
\Text(150,215)[t]{$H$}
\Text(20,305)[b]{$L$}
\Text(150,295)[b]{$L$}
\Text(95,260)[l]{$\Delta$}
\Text(85,210)[t]{\small{$\mu_{\Delta}$}}
\Text(85,305)[b]{\small$Y_{\Delta}$}
\LongArrow(85,185)(85,155)
\LongArrow(85,155)(85,185)
\ArrowLine(190,220)(255,220)
\ArrowLine(255,220)(255,300)
\ArrowLine(320,300)(255,300)
\DashArrowLine(190,300)(255,300){5}
\DashArrowLine(320,220)(255,220){5}
\Text(190,215)[t]{$L$}
\Text(320,215)[t]{$H$}
\Text(190,305)[b]{$H$}
\Text(320,295)[b]{$L$}
\Text(265,235)[l]{$N_1$}
\Text(265,285)[l]{$N_2$}
\Text(255,210)[t]{\small{$Y_{\nu}$}}
\Text(255,305)[b]{\small$\epsilon Y'_{\nu}$}
\LongArrow(255,185)(255,155)
\LongArrow(255,155)(255,185)
\ArrowLine(360,220)(425,220)
\ArrowLine(425,220)(425,260)
\ArrowLine(425,300)(425,260)
\ArrowLine(490,300)(425,300)
\DashArrowLine(360,300)(425,300){5}
\DashArrowLine(490,220)(425,220){5}
\Text(360,215)[t]{$L$}
\Text(490,215)[t]{$H$}
\Text(360,305)[b]{$H$}
\Text(490,305)[b]{$L$}
\Text(435,230)[l]{$N_1$}
\Text(435,250)[l]{$N_2$}
\Text(435,270)[l]{$N_2$}
\Text(435,290)[l]{$N_1$}
\Text(422,260)[l]{$\huge{\times}$}
\Text(415,260)[r]{$\mu$}
\Text(425,210)[t]{\small{$Y_{\nu}$}}
\Text(425,305)[b]{\small$Y_{\nu}$}
\LongArrow(425,185)(425,155)
\LongArrow(425,155)(425,185)
\end{picture}
\caption{\it Example of genuine one loop seesaws (lower row) and their equivalents (upper row): type~II one loop seesaw leading to type~II seesaw with suppressed Yukawa coupling (left), type~I one loop seesaw leading to linear seesaw (middle), and type~I on loop seesaw leading to inverse seesaw (right).
}
\label{Fig:LoopVsInvSS}
\end{center}
\end{figure}

In all theses models, the smallness of the lepton number violating couplings suggests that lepton number is a protected by a symmetry. However, one can seek other natural explanation for such small couplings, which can be the loop generation in the discussed models. Indeed, from Fig.~\ref{Fig:LoopVsInvSS}, the loops in the different diagrams lead to  effective realizations of the lepton number violating couplings, where lepton number is violated by the presence of Majorana masses in the loop and the smallness is explained by both heavy masses and loop factor. The linear seesaw can be generated from the type~I one loop seesaw, see middle column. Since both Yukawas enter neutrino mass, see \equ{InvSS_mass}, but LFV may depend on the unsuppressed coupling, see Eq.~(\ref{eq:TypeI_III_LFV}), it can be large.  As an interesting difference compared to the usual type~II seesaw, see left column, the Yukawa coupling must be loop suppressed which enters LFV, \ie, large LFV cannot be expected in this case.  Finally, we point out that a loop generation of $\mu$ would involve a self-energy
diagram, see right panel, that we do not consider in this paper (see {\it e.g.}, Ref.~\cite{Ma:2009gu}).

\section{Summary and conclusions}
\label{sec:conclusions}

Because the neutrinos are much lighter than the other fermions in the Standard Model, they are believed to communicate with a higher energy scale which causes this suppression. Whereas this energy scale is the GUT scale for the usual seesaw mechanisms, additional suppression mechanisms, such as small lepton number violating couplings, radiative generation of neutrino mass, or neutrino mass from a higher than $d=5$ operator, may lower this energy scale. The most interesting case may be an energy scale as low as the TeV scale, which may be testable at the LHC. While many possibilities to obtain such a low scale have been pointed out in  the literature, we have systematically studied the radiative generation of neutrino mass from scalars and fermions at the one loop order, and we have compared our results to the literature.
 We have listed the relevant new, fields and their interactions that are necessary to generate this operator at one loop systematically, and we have computed for each model the contribution to neutrino masses. 

Our observations can be summarized as follows, referring to the topologies in \figu{one-loopTopo}:
\begin{itemize}
\item
Topologies~1 and~3 lead to finite neutrino masses, with quite a number of possibilities for the field within the loops. A famous example for this category is the Zee model. In some cases, couplings leading to a tree level contribution to neutrino mass are involved, which, however, can be easily forbidden by (at least) a discrete $\mathbb{Z}_2$ symmetry. This category is probably the best studied one in the literature.
\item
 Topologies~5 and~6 and some of the possibilities of topology~4 lead to divergent one loop contributions to neutrino mass. These possibilities do not constitute new possibilities  themselves, since they just correspond to vertex corrections to the tree level seesaws -- which cannot be forbidden in these cases.
 Topology~2 does not lead to neutrino masses at all.
\item
 A class of three finite diagrams of topology~4 has been identified. While there are many possibilities for field insertions in general, the tree level seesaws cannot be genuinely circumvented as leading contribution to neutrino mass by a discrete symmetry. However, we have demonstrated that in very few cases promoting a fermion in the loop to a Majorana fermion and assuming lepton number conserving couplings, the tree level seesaw are not allowed. These cases have been identifed as extensions of the type~II seesaw, for which the Yukawa couplings are generated by a loop, and of the type~I or~III seesaw, 
which has turned out to be similar to a linear seesaw.
In these cases, internal scalar fields have to be prevented from taking a vev, which can easily achieved by  discrete symmetry. 
\end{itemize}
Our  results are summarized in \Tab~\ref{tab:summary}. It is noteworthy that in the one loop type~II seesaw, large LFV processes are suppressed compared to the usual type~II seesaw.

We conclude that we have studied all possible one-loop realizations 
of the dimension-5 Weinberg operator. 
While some of the possibilities generating neutrino
mass at one loop have been well studied in the literature, there is 
at least one more
category which deserves further attention. In particular, it can be
interpreted as natural extension of the tree level seesaws, and it can
be possibly promoted to higher loop orders.

\begin{table}[t]
\begin{center}
\begin{tabular}{|c||c|c|c|c|}
\hline
&Topology & Conditions & Large LFV & Fields at TeV\\
\hline
\hline
\multirow{4}{*}{Irreducible mechanisms} & T1-i & $\mathbb{Z}_2$  if tree-level mediators  & $\times$ & \checkmark\\
\cline{2-5}
& T1-ii & $\mathbb{Z}_2$ if tree-level mediators & $\times$ & \checkmark\\
\cline{2-5}
& T1-iii& $\mathbb{Z}_2$  if tree-level mediators& $\times$ & \checkmark\\
\cline{2-5}
& T3 & $\mathbb{Z}_2$  if tree-level mediators & $\times$ & \checkmark\\
\hline
\hline
\multirow{2}{*}{One loop seesaws} & T4-2-i & $\begin{array}{c}\mathbb{Z}_2,\,\rm{Majorana\, fermions\, in\, loop},\\ \rm{LN\, conserving\, couplings}\end{array}$ & $\times$ & \checkmark\\
\cline{2-5}
& T4-3-i & $\begin{array}{c}\mathbb{Z}_2,\,\rm{Majorana\, fermions\, in\, loop},\\ \rm{LN\, conserving\, couplings}\end{array}$ &  \checkmark & \checkmark\\
\hline
\end{tabular}
\end{center}
\caption{\it Summary of the topologies leading to leading order loop neutrino masses, and their potential phenomenology.}
 \label{tab:summary}
 \end{table}

\section*{Acknowledgements}

M.H. acknowledges support from the Spanish MICINN grants 
FPA2011-22975 and MULTIDARK CSD2009-00064 and 
by the Generalitat Valenciana grant Prometeo/2009/091 and the
EU~Network grant UNILHC PITN-GA-2009-237920. 
F. B. and W. W. acknowledge support from DFG grant WI 2639/4-1,
and W. W. from DFG grant WI 2639/3-1.


\appendix

\section{Possible field assignments and neutrino mass\label{App:numass}}

In this appendix, we show the possible field and hypercharge $\alpha$ ($\alpha$: integer) assignments corresponding to each diagram,  as well as the expression for the neutrino they generate after EWSB.

\begin{landscape}
\begin{table}[p]
\begin{center}
\begin{tabular}{|c|c|c|c|c|c|}
\hline
Diagram & $m_{\nu}/\langle H\rangle^2$ & \multicolumn{4}{|c|}{Fields} \\
\hline
\hline
\multirow{9}{*}{\includegraphics[width=4.5cm]{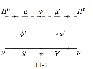} }  &  \multirow{9}{*}{$\mnuToneone$} & \FieldBoxOne{S}{S}{F}{S}{\alpha}{\alpha-1}{\alpha}{1+\alpha}\\
\hline 
\hline
\multirow{9}{*}{\includegraphics[width=4.5cm]{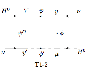}} &  \multirow{9}{*}{$\mnuTonetwo$} &  \FieldBoxTwo{F}{S}{S}{F}{\alpha}{1+\alpha}{\alpha}{1+\alpha} \\
\hline
\hline
\multirow{9}{*}{\includegraphics[width=4.5cm]{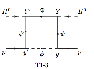}} &  \multirow{9}{*}{$\mnuTonethree$} &  \FieldBoxThree{F}{F}{S}{F}{\alpha}{1+\alpha}{\alpha}{\alpha-1}\\
\hline
\end{tabular}
\end{center}
\caption{\it Diagrams, neutrino masses, and possible fields for the T1 topology. In order to simplify the notation we used $M_a$ for $M_{\psi_a}$, $M'_{a'}$ for $M_{{\psi'}_{a'}}$ and $\mathcal{M}_A$ for $M_{\Psi_A}$.  }
\label{tab:massandchargeT1}
\end{table}
\end{landscape}

\begin{landscape}
\begin{table}[p]
\begin{center}
\begin{tabular}{|c|c|c|c|c|}
\hline
Diagram & $m_{\nu}/\langle H\rangle^2$ & \multicolumn{3}{|c|}{Fields} \\
\hline
\hline
\multirow{7}{*}{\includegraphics[width=4.5cm]{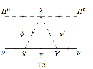} }  &  \multirow{7}{*}{$\mnuTthree$} & \FieldPyramid{F}{S}{F}{\alpha}{2+\alpha}{1+\alpha}\\
\hline 
\hline
\multirow{7}{*}{\includegraphics[width=4.5cm]{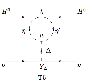} }  &  \multirow{7}{*}{$\mnuTfive$} & \FieldTadpoleFive{S}{S}{S}{-2}{\alpha}{2+\alpha}\\
\hline
\hline
\multirow{7}{*}{\includegraphics[width=4.5cm]{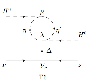} }  &  \multirow{7}{*}{$\mnuTsix$} & \FieldTadpoleSix{S}{S}{S}{-2}{\alpha}{1+\alpha}\\
\hline
\end{tabular}
\end{center}
\caption{\it Diagrams, neutrino masses, and possible fields for the  topologies T3, T5, T6. In order to simplify the notation we used $M_a$ for $M_{\psi_a}$.}
\label{tab:massandchargeT3T5T6}
\end{table}
\end{landscape}


\begin{landscape}
\begin{table}[p]
\begin{center}
\begin{tabular}{|c|c|c|c|c|c|}
\hline
Diagram & $m_{\nu}/\langle H\rangle^2$ & \multicolumn{4}{|c|}{Fields} \\
\hline
\hline
\multirow{7}{*}{\includegraphics[width=4.5cm]{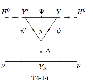} }  &  \multirow{7}{*}{$\mnuTfouronei$}  &\FieldTriangleTFourOnei{S}{F}{F}{F}{-2}{\alpha}{2+\alpha}{1+\alpha}\\
\hline 
\hline
\multirow{7}{*}{\includegraphics[width=4.5cm]{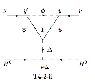} }  &  \multirow{7}{*}{$\mnuTfourtwoii$}  &\FieldTriangleTFourTwoii{S}{F}{F}{S}{-2}{\alpha}{\alpha-2}{\alpha-1}\\
\hline
\hline
\multirow{10}{*}{\includegraphics[width=4.5cm]{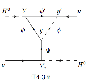} }  &  \multirow{10}{*}{$\mnuTfourthreeii$}  &\FieldTriangleTFourThreeii{F}{F}{S}{F}{0}{\alpha}{\alpha}{1+\alpha}\\
\hline
\end{tabular}
\end{center}
\caption{\it Diagrams, neutrino masses, and possible fields for the for the divergent diagrams of topology T4. In order to simplify the notation we used $M_a$ for $M_{\psi_a}$, $M'_{a'}$ for $M_{{\psi'}_{a'}}$ and $\mathcal{M}_A$ for $M_{\Psi_A}$. }
\label{tab:massandchargeT4div}
\end{table}
\end{landscape}


\begin{landscape}
\begin{table}[p]
\begin{center}
\begin{tabular}{|c|c|c|c|c|c|}
\hline
Diagram & $m_{\nu}/\langle H\rangle^2$ & \multicolumn{4}{|c|}{Fields} \\
\hline
\hline
\multirow{7}{*}{\includegraphics[width=4.5cm]{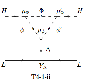} }  &  \multirow{7}{*}{$\mnuTfouroneii$}  &\FieldTriangleTFourOneii{S}{S}{S}{S}{-2}{\alpha}{2+\alpha}{1+\alpha}\\
\hline 
\hline
\multirow{7}{*}{\includegraphics[width=4.5cm]{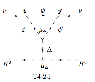} }  &  \multirow{7}{*}{$\mnuTfourtwoi$}  &\FieldTriangleTFourTwoi{S}{S}{S}{F}{2}{\alpha}{\alpha-2}{\alpha-1}\\
\hline
\hline
\multirow{10}{*}{\includegraphics[width=4.5cm]{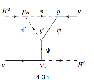} }  &  \multirow{10}{*}{$\mnuTfourthreei$}  &\FieldTriangleTFourThreei{F}{S}{S}{F}{0}{1+\alpha}{\alpha}{\alpha}\\
\hline
\end{tabular}
\end{center}
\caption{\it Diagrams, neutrino masses, and possible fields for the finite diagrams of topology T4. In order to simplify the notation we used $M_a$ for $M_{\psi_a}$. }
\label{tab:massandchargeT4fin}
\end{table}
\end{landscape}

The functions $I_n$ and $J_n$ used in Tables~\ref{tab:massandchargeT1}, \ref{tab:massandchargeT3T5T6},  \ref{tab:massandchargeT4div}, and \ref{tab:massandchargeT4fin} are defined as:
\begin{align}
I_{2} (M_{A}^{2}, M_{B}^{2})
\equiv&
\int \frac{{\rm d}^{d} k}{(2\pi)^{d} {\rm i}}
\frac{1}{
(k^{2} - M_{A}^{2})
(k^{2} - M_{B}^{2})
}\nonumber\\
=&
\frac{1}{(4\pi)^{2}}
 \left[
\frac{2}{\epsilon} 
-\gamma_E +1+\ln(4\pi)-\frac{M_A^2\ln(M_A^2)}{M_A^2-M_B^2}+\frac{M_B^2\ln(M_B^2)}{M_A^2-M_B^2}
\right],
\\
I_{3} (M_{A}^{2}, M_{B}^{2}, M_{C}^{2})
\equiv&
\int \frac{{\rm d}^{d} k}{(2\pi)^{d} {\rm i}}
\frac{1}{
(k^{2} - M_{A}^{2})
(k^{2} - M_{B}^{2})
(k^{2} - M_{C}^{2})
}
\nonumber 
\\
=&
-
\frac{1}{(4\pi)^{2}}
\left[
\frac{M_{A}^{2} \ln \frac{M_{C}^{2}}{M_{A}^{2}}}
{(M_{A}^{2} - M_{B}^{2})(M_{A}^{2} - M_{C}^{2})}
+
\frac{M_{B}^{2} \ln \frac{M_{C}^{2}}{M_{B}^{2}}}
{(M_{B}^{2} - M_{A}^{2})(M_{B}^{2} - M_{C}^{2})}
\right],
\\
I_{4} (M_{A}^{2}, M_{B}^{2}, M_{C}^{2}, M_{D}^{2})
\equiv&
\int \frac{{\rm d}^{d} k}{(2\pi)^{d} {\rm i}}
\frac{1}{
(k^{2} - M_{A}^{2})
(k^{2} - M_{B}^{2})
(k^{2} - M_{C}^{2})
(k^{2} - M_{D}^{2})
}
\nonumber 
\\
=&
-
\frac{1}{(4\pi)^{2}}
\left[
\frac{M_{A}^{2} \ln \frac{M_{D}^{2}}{M_{A}^{2}}}
{
(M_{A}^{2} - M_{B}^{2})
(M_{A}^{2} - M_{C}^{2})
(M_{A}^{2} - M_{D}^{2})
}
\right.
\nonumber 
\\
&
\hspace{1.5cm}
+
\frac{M_{B}^{2} \ln \frac{M_{D}^{2}}{M_{B}^{2}}}
{
(M_{B}^{2} - M_{A}^{2})
(M_{B}^{2} - M_{C}^{2})
(M_{B}^{2} - M_{D}^{2})
}\nonumber
\\
&
\hspace{1.5cm}
+
\left.
\frac{M_{C}^{2} \ln \frac{M_{D}^{2}}{M_{C}^{2}}}
{
(M_{C}^{2} - M_{A}^{2})
(M_{C}^{2} - M_{B}^{2})
(M_{C}^{2} - M_{D}^{2})
}
\right],
\\
J_{3} (M_{A}^{2}, M_{B}^{2}, M_{C}^{2})
\equiv&
\int \frac{{\rm d}^{d} k}{(2\pi)^{d} {\rm i}}
\frac{k^{2}}{
(k^{2} - M_{A}^{2})
(k^{2} - M_{B}^{2})
(k^{2} - M_{C}^{2})
}
\nonumber
\\
=&
\frac{1}{(4\pi)^{2}}
\left[
 \frac{2}{\epsilon}
-
\gamma_{E}
+
1
+
\ln(4\pi)
-
\ln M_{C}^{2}
\right.
\nonumber 
\\
&\hspace{1cm}\left.
+
\frac{M_A^4 \ln \frac{M_C^2}{M_{A}^{2}}}
{(M_A^2-M_B^2)(M_A^2-M_C^2)}
+
\frac{M_B^4 \ln \frac{M_{C}^{2}}{M_B^2}}
{(M_B^2-M_A^2)(M_B^2-M_C^2)}
\right]
,
\\
J_{4} (M_{A}^{2}, M_{B}^{2}, M_{C}^{2}, M_{D}^{2})
\equiv&
\int \frac{{\rm d}^{d} k}{(2\pi)^{d} {\rm i}}
\frac{k^{2}}{
(k^{2} - M_{A}^{2})
(k^{2} - M_{B}^{2})
(k^{2} - M_{C}^{2})
(k^{2} - M_{D}^{2})
}
\nonumber 
\\
=&
-
\frac{1}{(4\pi)^{2}}
\left[
\frac{M_{A}^{4} \ln \frac{M_{D}^{2}}{M_{A}^{2}}}
{
(M_{A}^{2} - M_{B}^{2})
(M_{A}^{2} - M_{C}^{2})
(M_{A}^{2} - M_{D}^{2})
}
\right.
\nonumber 
\\
&
\hspace{1.5cm}
+
\frac{M_{B}^{4} \ln \frac{M_{D}^{2}}{M_{B}^{2}}}
{
(M_{B}^{2} - M_{A}^{2})
(M_{B}^{2} - M_{C}^{2})
(M_{B}^{2} - M_{D}^{2})
}
\nonumber
\\
&
\hspace{1.5cm}
\left.
+
\frac{M_{C}^{4} \ln \frac{M_{D}^{2}}{M_{C}^{2}}}
{
(M_{C}^{2} - M_{A}^{2})
(M_{C}^{2} - M_{B}^{2})
(M_{C}^{2} - M_{D}^{2})
}
\right].
\end{align}
Note that the functions $I_2$ and $J_3$ imply divergent contributions, whereas the other
functions correspond to finite contributions.


\bibliography{references_oneloop}
\bibliographystyle{h-physrev5.bst}

\end{document}